\documentclass[twocolumn]{aastex63}

\usepackage{amsmath}
\usepackage{enumerate}
\input{colordvi}
\listofchanges
\hypersetup{
    unicode=false,                  
    pdftoolbar=true,                
    pdfmenubar=true,                
    pdffitwindow=true,              
    pdfstartview={FitH},            
    pdftitle={J1109+0754},          
    pdfauthor={m.mardini},          
    pdfsubject={Astronomy},         
    pdfcreator={dvipdf},            
    pdfproducer={dvipdf},           
    pdfkeywords={metal-poor stars}, 
    pdfnewwindow=true,              
    colorlinks=true,                
    linkcolor=red,                  
    citecolor=blue,                 
    filecolor=magenta,              
    urlcolor=cyan,                  
    breaklinks=true,
    linktocpage
}

\newcommand{\rapo}{r_\ensuremath{\mathrm{apo}}}
\newcommand{\rperi}{r_\ensuremath{\mathrm{peri}}}
\newcommand{\zmax}{Z_\ensuremath{\mathrm{max}}}
\def\vector#1{\mbox{\boldmath $#1$}}
\def\be{\begin{equation}}
\def\ee{\end{equation}}

\def\ms{\ifmmode {\rm M_{\odot}} \else ${\rm M_{\odot}}$\fi}

\newcommand{\abund}[2]{\ensuremath{[\mathrm{#1}/\mathrm{#2}]}}

\newcommand{\cemp}{{CEMP}}
\newcommand{\teff}{\ensuremath{T_\mathrm{eff}}}
\newcommand{\logg}{\ensuremath{\log\,g}}

\newcommand{\A}[1]{\ensuremath{A(\mathrm{#1})}}

\newcommand{\K}{\ensuremath{\mathrm{K}}}

\definecolor{BrickRed}{HTML}{F26035}

\def\PGRAPE{$\varphi-$GRAPE }

\definecolor{Ultramarine}{HTML}{120A8F}
\definecolor{Deepskyblue}{HTML}{00BFFF}

\submitjournal{ApJ}

\shorttitle{LAMOST J1109+0754}
\shortauthors{Mardini et al.}

\begin{document} 

\correspondingauthor{Gang Zhao}
\email{gzhao@nao.cas.cn}

\title{Cosmological Insights into the Early Accretion of \textit{r}-Process-Enhanced stars. I. A Comprehensive Chemo-dynamical Analysis of LAMOST J1109+0754}

\author[0000-0001-9178-3992]{Mohammad K.\ Mardini}
\affiliation{Key Lab of Optical Astronomy, National Astronomical Observatories, Chinese Academy of Sciences, Beijing 100101, China}
\affiliation{Institute of Space Sciences, Shandong University, Weihai 264209, China}
\author[0000-0003-4479-1265]{Vinicius M.\ Placco }
\affiliation{NSF's Optical-Infrared Astronomy Research Laboratory, Tucson, AZ 85719, USA}
\author[0000-0003-3518-5183]{Yohai Meiron}
\affiliation{Department of Astronomy and Astrophysics, University of Toronto, 50 St. George Street, Toronto, ON M5S\,3H4, Canada}
\author[0000-0002-6961-8170]{Marina Ishchenko}
\affiliation{Main Astronomical Observatory, National Academy of Sciences of Ukraine, 27 Akademika Zabolotnoho St., 03143 Kyiv, Ukraine}
\author[0000-0002-1638-3080]{Branislav Avramov}
\affiliation{Astronomisches Rechen Institut - Zentrum f\"ur Astronomie der Universit\"at Heidelberg, M\"onchhofstrasse 12-14, D-69120 Heidelberg, Germany}
\author[0000-0002-9293-821X]{Matteo Mazzarini}
\affiliation{Astronomisches Rechen Institut - Zentrum f\"ur Astronomie der Universit\"at Heidelberg, M\"onchhofstrasse 12-14, D-69120 Heidelberg, Germany}
\author[0000-0003-4176-152X]{Peter Berczik}
\affiliation{Main Astronomical Observatory, National Academy of Sciences of Ukraine, 27 Akademika Zabolotnoho St., 03143 Kyiv, Ukraine}
\affiliation{Astronomisches Rechen Institut - Zentrum f\"ur Astronomie der Universit\"at Heidelberg, M\"onchhofstrasse 12-14, D-69120 Heidelberg, Germany}
\affiliation{National Astronomical Observatories and Key Laboratory of Computational Astrophysics, Chinese Academy of Sciences, 20A Datun Rd., Chaoyang District, Beijing 100101, China}
\author[0000-0002-3987-0519]{Manuel Arca Sedda}
\affiliation{Astronomisches Rechen Institut - Zentrum f\"ur Astronomie der Universit\"at Heidelberg, M\"onchhofstrasse 12-14, D-69120 Heidelberg, Germany}
\author[0000-0003-4573-6233]{Timothy C. Beers}
\affiliation{Department of Physics, University of Notre Dame, Notre Dame, IN 46556, USA}
\affiliation{JINA Center for the Evolution of the Elements (JINA-CEE), USA}
\author[0000-0002-2139-7145]{Anna Frebel}
\affiliation{Department of Physics and Kavli Institute for Astrophysics and Space Research, Massachusetts Institute of Technology, Cambridge, MA 02139, USA}
\affiliation{JINA Center for the Evolution of the Elements (JINA-CEE), USA}
\author[0000-0002-1558-1472]{Ali Taani}
\affiliation{Physics Department, Faculty of Science, Al-Balqa Applied University,Jordan}
\author[0000-0001-8903-7026]{Martina Donnari}
\affiliation{Max-Planck-Institut f\"ur Astronomie, K\"onigstuhl 17, 69117 Heidelberg, Germany}
\author[0000-0002-1422-211X]{Mashhoor A. Al-Wardat}
\affiliation{Department of Applied Physics and Astronomy, University of Sharjah, Sharjah, United Arab Emirates}
\affiliation{Sharjah Academy for Astronomy, Space Sciences and Technology, University of Sharjah, Sharjah, United Arab Emirates}
\author[0000-0002-8980-945X]{Gang Zhao}
\affiliation{Key Lab of Optical Astronomy, National Astronomical Observatories, Chinese Academy of Sciences, Beijing 100101, China}
\affiliation{Institute of Space Sciences, Shandong University, Weihai 264209, China}

\begin{abstract}
This study presents a comprehensive chemo-dynamical analysis of LAMOST J1109+0754, a bright (V = 12.8), extremely metal-poor (\abund{Fe}{H} = $-3.17$) star, with a strong \textit{r}-process enhancement (\abund{Eu}{Fe} = +0.94 $\pm$ 0.12). Our results are based on the 7-D measurements supplied by $Gaia$ and the chemical composition derived from a high-resolution ($R\sim 110,000$), high signal-to-noise ratio ($S/N \sim 60)$ optical spectrum obtained by the 2.4\,m Automated Planet Finder Telescope at Lick Observatory. 
We obtain chemical abundances of 31 elements (from lithium to thorium). The abundance ratios (\abund{X}{Fe}) of the light-elements (Z $\leqslant 30$) suggest a massive Population\,III progenitor in the 13.4-29.5\,M$_\odot$ mass range. The heavy-element ($30 <$ Z $\leqslant 90$) abundance pattern of J1109+075 agrees extremely well with the scaled-Solar \textit{r}-process signature. We have developed a novel approach to trace the kinematic history and orbital evolution of J1109+0754 with a c\textbf{O}smologically de\textbf{RI}ved tim\textbf{E}-varyi\textbf{N}g Galactic po\textbf{T}ential (the ORIENT) constructed from snapshots of a simulated Milky-Way analog taken from the \texttt{Illustris-TNG} simulation. The orbital evolution within this Milky Way-like galaxy, along with the chemical-abundance pattern implies that J1109+0754 likely originated in a low-mass dwarf galaxy located $\sim$ 60\,kpc from the center of the Galaxy, which was accreted $\sim$ 6 - 7\,Gyr ago, and that the star now belongs to the outer-halo population.
\end{abstract}


\section{Introduction} \label{sec:intro}

Since the pioneering work of \citet{1957RvMP...29..547B} and \citet{1957PASP...69..201C}, numerous studies have focused attention on the astrophysical site(s) of the rapid neutron-capture process (\textit{r}-process). Although full understanding has not yet been achieved, several promising mechanisms have been proposed, including: i) the innermost ejecta of regular core-collapse supernovae \citep[e.g.,][]{1974PThPh..51..726S, 1994A&A...286..841W, 2010ApJ...712.1359F,2015PhRvD..92j5020M}, ii) the outer layers of supernova explosions \citep[e.g.,][]{1979A&A....74..175T, 1983ApJ...265..429C, 2007AstL...33..385N, 2014JPhG...41d4002Q}, iii) magneto-rotational jet-driven supernovae \citep[e.g.,][]{1985ApJ...291L..11S, 2008ApJ...680.1350F, 2015ApJ...810..109N, 2018JPhG...45h4001O}, iv) neutron star mergers (NSMs) \citep[e.g.,][]{1982ApL....22..143S, 2000A&A...360..171R, 2015ApJ...808...30E, 2017ARNPS..67..253T}, and v) collapsars \citep{2019Natur.569..241S}. 

Observationally, the advanced LIGO-Virgo detectors collected the first gravitational wave signature from the merger of a binary neutron star system \citep[GW170817;][]{PhysRevLett.119.161101}, 
with subsequent electromagnetic follow-up (photometric and spectroscopic) observations of its associated kilonova
(SSS17a; e.g., \citealt{2017Sci...358.1570D,2017Sci...358.1583K,2017Sci...358.1574S}), \textbf{(Added: providing a prime example of multi-messenger astronomy)} \citep[e.g.,][]{Abbott_2017, Goldstein_2017,  Savchenko_2017, Soares_Santos_2017}. These kilonova observations provided strong evidence for the existence of at least one site for the astrophysical operation of the $r$-process, namely NSMs \citep[e.g.,][]{2017ApJ...848L..17C, 2017ApJ...836..230C}. Even before this detection, support for NSMs as a potential major source of $r$-process elements was found with the discovery of the ultra-faint dwarf galaxy Reticulum\,II \citep{2016Natur.531..610J} that contains almost exclusively $r$-process-enhanced metal-poor stars (see more details below) that appear to have originated in this system from gas that had been enriched by a prior, prolific $r$-process event that polluted this galaxy. Models of the yields suggested this event to have been a NSM, but other sources might also have contributed.

Interestingly, \citet{Placco_2020} recently analyzed the moderately $r$-process and CNO-enhanced star RAVE J1830$-$4555, whose neutron-capture abundance pattern matches both the fast ejecta yields of a NSM as well as the yields of a rotating massive star experiencing an $r$-process event during its explosion. Magneto-rotational supernovae have been suggested a viable astrophysical environments for the main \textit{r}-process to operate \citep[e.g.,][]{2017ApJ...836L..21N, 2018MNRAS.477.2366H,2018JPhG...45h4001O, 2019ApJ...875..106C}, but more discriminating model predictions are needed to make progress, in addition to more observations, to fully investigate these (multiple) progenitor sources. This is supported by the overall observed levels of \abund{Eu}{Fe}\footnote{[X/Y] = log(N$_{X}$/N$_{Y}$)$_{\star} - $log(N$_{X}$/N$_{Y}$)$_{\odot}$, where N is the number density of atoms of elements X and Y in the star ($\star$) and the Sun ($\odot$), respectively.} in the body of data of, e.g., metal-poor stars, which suggests that more than just one source is responsible for the $r$-process inventory of the Universe. 

In particular, constraints can be uniquely obtained from individual Galactic halo stars with enhancements in \textit{r}-process elements -- the so-called \textit{r}-process-enhanced (RPE) stars --  to provide novel insights into this long-standing issue \citep[for a selected list see, e.g.,][and references therein]{1996ApJ...467..819S, 2002A&A...387..560H, 2012A&A...545A..31H, 2017ApJ...844...18P, 2018MNRAS.481.1028H, 2018ApJ...865..129R, 2018ApJ...854L..20S, 2018ApJ...868..110S,2019ApJ...874..148S,Placco_2020}.

Substantial recent efforts have been underway to increase the numbers of the known RPE metal-poor stars \citep{2004A&A...428.1027C, 2005A&A...439..129B, Mardini_2019a}, including that of the \textit{R}-Process Alliance (RPA) \citep{2018ApJ...858...92H, 2018ApJ...868..110S, 2020arXiv200607731E, 2020arXiv200700749H} which have recently identified a total of 72 new \textit{r}-II and 232 new \textit{r}-I stars\footnote{\textbf{(Added: \textit{r}-II stars are defined as \abund{Eu}{Fe}$>+1.0$ and \abund{Ba}{Eu} $<0$, While \textit{r}-I stars are defined as $+0.3 <$ \abund{Eu}{Fe} $\leqslant +1.0$ and \abund{Ba}{Eu} $<0$.)}}. This has increased the number of RPE stars known to 141 \textit{r}-II and 345 \textit{r}-I stars. Here we report on a detailed analysis of \object{LAMOST J110901.22+075441.8} (hereafter \object[LAMOST J110901.22+075441.8]{J1109+0754}), an $r$-II star with strong carbon enhancement, originally identified by \citep{2015RAA....15.1264L}. J1109+0754 thus adds to the sample of well-studied RPE stars.

It is now widely recognized that the halo of the Milky Way experienced mergers with small dwarf galaxies, and grew hierarchically as a function of time \citep[e.g.,][]{1978ApJ...225..357S,10.1093/mnras/183.3.341, 1985ApJ...292..371D}. Some of these galaxies have survived, some experienced strong structural distortions, and some have been fully disrupted. Since the discovery of Reticulum\,II \citep{2016Natur.531..610J}, it has become clear that at least some of the halo RPE stars must have originated in small satellite dwarf galaxies \citep{2019ApJ...871..247B} before their eventual accretion into the Galactic halo. Therefore, combining chemical compositions of RPE stars with the results from kinematic analyses and/or results from cosmological simulations can help to assess the cosmic origin of these stars, in addition to learning about the formation and evolution of the Milky Way (e.g., \citealt{Roederer_2018, Mardini_2019a}, and Gudin et al. 2020, in prep). 
The orbital integrations of halo stars reported in the literature, including those of RPE stars, \citep[e.g.,][]{Roederer_2018, Mardini_2019a} are usually determined with a {\it fixed} Galactic potential (e.g., \texttt{MWPotential2014};  \citealt{2015ApJS..216...29B}). This fixed Galactic potential provides a snapshot view of the present-day dynamical parameters of, e.g., RPE stars, but in order to gain detailed insights into the stars' orbital histories in a more realistic way, a {\it time-varying} Galactic potential should be considered. In this study, we explore a time-varying Galactic potential, based on a simulated Milky-Way analog  extracted from the \texttt{Illustris-TNG} simulation \citep{2015MNRAS.449...49R,2018MNRAS.480.5113M, 2018MNRAS.477.1206N,  2018MNRAS.475..624N, 2018MNRAS.475..648P, 2018MNRAS.475..676S, 2019MNRAS.490.3234N, 2019ComAC...6....2N, 2019MNRAS.490.3196P}. This allows us to gain a more complete picture of the orbital evolution of J1109+0754. 

Cosmological simulations can nowadays be carried out with a sufficiently large number of tracer particles such that Milky Way-sized halos can be well-resolved, including both their baryonic components (gas and stars). By identifying halos in the simulation box that are representative of the Milky Way, we aim at mapping the evolution of the orbit of J1109+0754 to learn about its possible origin scenario.

This paper is organized as follows. We describe the observational data in Section~\ref{sec:observations}. Section~\ref{sec:parameters} presents the determinations of stellar parameters. The chemical abundances are addressed in Section~\ref{sec:abundances}. The possible pathways that may have led to the formation of J1109+0754 are described in Section~\ref{sec:discussion}. The kinematic signature and orbital properties of J1109+0754 are discussed in Section~\ref{sec:kinematics}. Our conclusions are presented in Section~\ref{sec:conclusions}.

\section{Observations}\label{sec:observations}

\begin{deluxetable*}{lcccl}
\tablenum{1}
\tablecaption{Basic Data for LAMOST J1109+0754} \label{tab:observations}
\tablewidth{0pt}
\tablecolumns{5}
\tabletypesize{\small}
\tablehead{\colhead{Quantity} &\colhead{Symbol} &\colhead{Value} &\colhead{Units} &\colhead{Reference}}
\startdata
Right ascension           & $\alpha$ (J2000)    & 11:09:01.22            & hh:mm:ss.ss   & Simbad \\
Declination               & $\delta$ (J2000)    & +07:54:41.8            & dd:mm:ss.s    & Simbad \\
Galactic longitude        & $\ell$              & 246.5691               & degrees       & This Study \\
Galactic latitude         & $b$                 & 59.0610                & degrees       & This Study \\
Parallax                  & $\varpi$            & 0.0717 $\pm$ 0.0442    & mas           & \citet{refId0} \\
Distance                  & $D$                 & 4.91$^{+0.85}_{-0.69}$ & kpc            & \citet{2018AJ....156...58B} \\
Proper motion ($\alpha$)  & PMRA                & 2.009 $\pm$ 0.0748     & mas yr$^{-1}$ & \citet{refId0} \\
Proper motion ($\delta$)  & PMDec               & $-$9.155 $\pm$ 0.0652  & mas yr$^{-1}$ & \citet{refId0} \\
Mass                      & M                   & 0.75 $\pm$ 0.20        & $M_{\odot}$         & assumed \\
$B$ magnitude             & $B$                 & 13.361$\pm$ 0.009      & mag           & \citet{2016yCat.2336....0H} \\
$V$ magnitude             & $V$                 & 12.403$\pm$ 0.016      & mag           & \citet{2016yCat.2336....0H} \\
$J$ magnitude             & $J$                 & 10.443$\pm$ 0.024      & mag           & \citet{2006AJ....131.1163S} \\
$K$ magnitude             & $K$                 & 9.783$\pm$ 0.027       & mag           & \citet{2006AJ....131.1163S} \\
Color excess              & $E(B-V)$            & 0.0251$\pm$ 0.0005     & mag           & \citet{1984ApJ...278L...1N}\tablenotemark{a} \\
Bolometric correction     & $BC_{V}$            & $-$0.45 $\pm$ 0.07     & mag           & this study, based on \citet{Alonso} \\
Effective temperature     & \teff              	& 4633 $\pm$ 150         & K             & this study, based on \citet{2013ApJ...769...57F}  \\
Log of surface gravity    & \logg               & 0.96 $\pm$ 0.30        & dex         & this study, based on \citet{2013ApJ...769...57F} \\
Microturbulent velocity   & $v_{\rm t}$         & 2.20 $\pm$ 0.30        & km~s$^{\rm -1}$        & this study, based on \citet{2013ApJ...769...57F}\\
Metallicitiy              & [Fe/H]              & $-$3.17 $\pm$ 0.09     & \nodata           &  this study, based on \citet{2013ApJ...769...57F}\\
Radial velocity           & RV                  & $-$100.2 $\pm$ 1.02    & km~s$^{\rm -1}$        &  APF (MJD: 57132.229) \\
			  & RV                  & $-$98.64 $\pm$ 0.52    & km~s$^{\rm -1}$        &  \citet{refId0} \\
			  & RV                  & $-$99.05 $\pm$ 0.40    & km~s$^{\rm -1}$        &  SUBARU \citep[MJD: 56786.294,][]{2015RAA....15.1264L}\\
			  & RV                  & $-$102.9    $\pm$ 1.41    & km~s$^{\rm -1}$     & LAMOST (MJD: 581449.000, private communication) \\
Natal carbon abundance              & [C/Fe]             & +0.66  $\pm 0.27$                               &   \nodata                       & this study, based on \citet{2014ApJ...797...21P}  \\			
\enddata 
\tablenotetext{a}{https://irsa.ipac.caltech.edu/applications/DUST/}     
\tablenotetext{b}{Determined in the same manner presented in \citet{Mardini_2019a}}     
\end{deluxetable*}

The metal deficiency of J1109+0754 was firstly reported in the third data release (DR3\footnote{http://dr3.lamost.org}) of the Large Sky Area Multi-Object Fiber Spectroscopic Telescope (LAMOST) survey \citep{2006ChJAA...6..265Z, 2012RAA....12..723Z, 2012RAA....12.1197C}. This relatively bright (V = 12.8) K-giant star was originally followed-up with high-resolution spectroscopy on 2014, May 9, using the SUBARU telescope and the echelle High Dispersion Spectrograph (HDS, \citealt{10.1093/pasj/54.6.855}). The analysis of this high-resolution spectrum confirms the star's extremely low metallicity and strong enhancement in \textit{r}-process elements \citep{2015RAA....15.1264L}.

On 2015, May 14, J1109+0754 was also observed with the Automated Planet Finder Telescope (APF) (we refer the reader to more details about the target selection and the overall scientific goals in \citealt{Mardini_2019a}). The observing setup yielded a spectral resolving power of $R \sim$ 110,000. Note that our initial sample was selected according to the stars' corresponding Lick indices, as part of  a sample of 20 stars observed with APF. Data for 13 of those stars had sufficient S/N to allow for a reliable analyses, including J1109+0754. The results of the remaining 12 newly discovered stars were reported in \citet{Mardini_2019b,Mardini_2019a}. Due to the reduced wavelength coverage of the SUBARU spectrum, relatively few neutron-capture elements could be detected. Therefore, we decided to carry out a detailed analysis of the APF spectrum of J1109+0754 to complete its abundance assay, and link it to the star's kinematics. Table~\ref{tab:observations} lists the basic data for J1109+0754.

We used IRAF \citep{1986SPIE..627..733T, 1993ASPC...52..173T} to carry out a standard echelle data reduction (including bias subtraction, cosmic-ray removal, and wavelength calibration, etc). Our final APF spectrum covers a wide wavelength range ($\sim$ $3730$-$9989$\,{\AA}) and has a fairly good signal-to-noise ratio (S/N \textbf{(Added: per pixel)} $\sim$ 60 at 4500\,{\AA}). We measured the radial velocity (RV) of J1109+0754 in the same way described in \citet{Mardini_2019b}. We employed a synthesized template for the cross-correlation against the final reduced spectrum of J1109+0754; using the Mg I line triplet (at $\sim 5160-5190$\,{\AA}). This yielded RV = 82.39 $\pm 0.8$ km s$^{-1}$. In addition, J1109+0745 has some other radial velocity measurements in the literature (see Table~\ref{tab:observations}). \textbf{(Added: These measurements do not suggest the presence of an unseen binary companion, however, they do not exclude the possibility of a long-period binary)}.

\begin{deluxetable}{lccrrr}
\tablenum{2}
\tablecaption{Lines, Atomic Data, EWs, and Individual Abundances \label{tab:EWs}}
\tablewidth{0pt}
\tabletypesize{\scriptsize}
\tablehead{\colhead{Species} & \colhead{Wavelength} &\colhead{E.P.} &\colhead{$\log gf$}
 &\colhead{EW}  &\colhead{$\log\epsilon$  (X)} \\
\colhead{} & \colhead{(\AA)} &\colhead{(eV)} & \colhead{} &   \colhead{(m\AA)} &  \colhead{}}
\startdata
Li~\textsc{i}&  6707.749&  0.000&  $-$0.804&Syn&   $<-$0.02\\
C(CH)         &4214.000 &\nodata&   \nodata& Syn& 5.16 \\
O~\textsc{i}&  6300.300&  0.000&  $-$9.820&Syn&   6.57\\
Na~\textsc{i}&  5889.951&  0.000&  0.120&Syn&   3.37\\
Na~\textsc{i}&  5895.924&  0.000& $-$0.180&Syn&   3.26\\
Mg~\textsc{i}&  4167.270&  4.350& $-$0.710&  26.73&   4.89\\
Mg~\textsc{i}&  4571.100&  0.000& $-$5.690&  55.76&   5.06\\
Mg~\textsc{i}&  4702.990&  4.330& $-$0.380&  40.76&   4.72\\
Mg~\textsc{i}&  5172.684&  2.710& $-$0.400&Syn&   5.13\\
Mg~\textsc{i}&  5183.604&  2.715& $-$0.180&Syn&   5.13\\
Mg~\textsc{i}&  5528.400&  4.340& $-$0.500&  47.11&   4.86\\
Ca~\textsc{i}&  4283.010&  1.890& $-$0.220&  36.58&   3.55\\
Ca~\textsc{i}&  4318.650&  1.890& $-$0.210&  34.42&   3.49\\
Ca~\textsc{i}&  4425.440&  1.880& $-$0.360&  23.97&   3.38\\
Ca~\textsc{i}&  4435.690&  1.890& $-$0.520&  49.12&   4.04\\
Ca~\textsc{i}&  4454.780&  1.900&  0.260&  55.33&   3.38\\
Ca~\textsc{i}&  4455.890&  1.900& $-$0.530&  25.32&   3.60\\
Ca~\textsc{i}&  5265.560&  2.520& $-$0.260&  16.32&   3.72\\
Ca~\textsc{i}&  5588.760&  2.520&  0.210&  27.50&   3.51\\
Ca~\textsc{i}&  5594.470&  2.520&  0.100&  17.49&   3.37\\
Ca~\textsc{i}&  5598.490&  2.520& $-$0.090&  14.67&   3.47\\
Ca~\textsc{i}&  5857.450&  2.930&  0.230&  11.27&   3.49\\
Ca~\textsc{i}&  6102.720&  1.880& $-$0.790&  22.20&   3.59\\
Ca~\textsc{i}&  6122.220&  1.890& $-$0.310&  38.10&   3.44\\
Ca~\textsc{i}&  6162.170&  1.900& $-$0.090&  52.97&   3.47\\
Ca~\textsc{i}&  6439.070&  2.520&  0.470&  36.24&   3.36\\
Sc~\textsc{ii}&  4314.080&  0.620& $-$0.100&  81.98&   0.07\\
Sc~\textsc{ii}&  4325.000&  0.600& $-$0.440&  64.78&   0.05\\
Sc~\textsc{ii}&  4400.390&  0.610& $-$0.540&  49.85&  $-$0.12\\
\enddata
\tablecomments{Table \ref{tab:EWs} is published in its entirety in the machine-readable format. A portion is shown here for guidance regarding its form and content.}
\end{deluxetable}
\section{Determinations of Stellar Parameters}\label{sec:parameters}

We employed the \texttt{TAME} code \citep[for more details, see][]{2015ascl.soft03003K} to measure the equivalent widths (hereafter EWs) for 209 unblended lines of light elements (Z $\leqslant 30$), with the exception of Li, C, O, Na, and the two strong Mg line at 5172\,{\AA} and 5183\,{\AA} whose abundances we obtained from spectrum synthesis. We then applied the most recent version of the LTE stellar analysis code MOOG \citep{1973PhDT.......180S,2011AJ....141..175S} and one-dimensional $\alpha$-enhanced ([$\alpha$/Fe] = +0.4) atmospheric models \citep{2003IAUS..210P.A20C} to derive individual abundances from these lines. Table~\ref{tab:EWs} lists the atomic data used in this work, the measured EWs, and the derived individual abundances, including those from spectrum synthesis.

We employed 99 \ion{Fe}{1} and 12 \ion{Fe}{2} lines, which were used to spectroscopically determine the stellar parameters of J1109+0754. We reduced the slope of the derived individual line abundances of the \ion{Fe}{1} lines as a function of their excitation potentials (see Table \ref{tab:EWs}) to its minimum value in order to determine the effective temperature ($\teff$). By fixing the value of $\teff$ and also removing the trend between the individual abundances of \ion{Fe}{1} lines and reduced equivalent width, we determine the microturbulent velocity ($v_{\rm t}$). The surface gravity ($\logg$) was obtained by matching the average abundance of of the \ion{Fe}{1} and \ion{Fe}{2} lines. This procedure yielded $\teff$ = 4403\,K, $\logg$ = 0.11, and $v_{\rm t}$ = 2.87\,km\,s$^{\rm -1}$. These stellar parameters were used as inputs for the empirical calibration described in \citet{2013ApJ...769...57F} to adjust $\teff$ to the  photometric scale. This yielded final stellar parameters of $\teff$ = 4633\,K, $\logg$ = 0.96, \abund{Fe}{H}= $-3.17$, and $v_{\rm t}$ = 2.20\,km\,s$^{\rm -1}$, listed in Table~\ref{tab:observations}. We adopt these parameters in our remaining analysis.

We also employed the empirical metallicity-dependent color-$\teff$ relation presented by \citet{Alonso} to calculate the photometric $\teff$ of J1109+0754. We adopted total Galactic reddening along the line of sight to J1109+0754 of $E(B-V) = 0.025 \pm 0.001$ \citep{1984ApJ...278L...1N}. We generated 10,000 sets of parameter estimates, by re-sampling each input photometric information (the $B$, $V$, $J$, $K$ magnitudes, $E(B-V)$), along with \abund{Fe}{H}). We adopted the median of each calculation as the final results, and the 16th percentile and 84th percentiles (the subscript and superscript, respectively) as the uncertainties. These calculations yielded \textbf{(Added: $\teff (V-J) = 4575^{33}_{31}$\,K and $\teff (V-K) = 4521^{26 }_{26}$\,K.)}  These results are consistent, within $1 \sigma$, with the final $\teff$ of 4633\,K derived from the spectroscopic method.

\begin{deluxetable}{lrrrrrr}
\tablenum{3}
\tablecaption{Abundances of J1109+0754\label{tab:abundances}}
\tablewidth{0pt}
\tablecolumns{7}
\tabletypesize{\scriptsize}
\tablehead{ \colhead{Species} & \colhead{$\log\varepsilon$ (X)} & \colhead{[X/H]} &
 \colhead{[X/Fe]} & \colhead{$\sigma$ (dex)} & \colhead{$N_{\rm lines}$} &  \colhead{$\log\varepsilon_{\odot}$\tablenotemark{a}} \\}
\startdata
Li~\textsc{i}  & $< -0.02$ & $<-$1.07 &$<$2.10 &\nodata & 1  & 1.05 \\
C (CH)          & +5.16 &  $-$3.27 &    $-$0.10 &  0.10 &\nodata & 8.43 \\
O~\textsc{i}   & +6.57  & $-$2.12 & +1.05 & 0.10 & 1 & 8.69 \\
Na~\textsc{i}  &  +3.32 & $-$2.92 & +0.24 & 0.08 & 2           & 6.24 \\
Mg~\textsc{i}  &+4.96 & $-$2.64 & +0.53 & 0.17 & 6 & 7.60 \\
Ca~\textsc{i}  &+3.52 & $-$2.82 & +0.35 & 0.17 & 15 & 6.34 \\
Sc~\textsc{ii} &+0.01 & $-$3.14 & +0.03 & 0.08 & 6 & 3.15 \\
Ti~\textsc{i}  &+2.08 & $-$2.87 & +0.30 & 0.17 & 18 & 4.95 \\
Ti~\textsc{ii} &+2.08 & $-$2.87 & +0.30 & 0.17 & 35 & 4.95 \\
V~\textsc{i}  &+0.69 & $-$3.24 & $-$0.07 & 0.12 & 2 & 3.93 \\
Cr~\textsc{i}  &+2.04 & $-$3.60 & $-$0.43 & 0.17 & 6 & 5.64 \\
Mn~\textsc{i}  &+1.68 & $-$3.75 & $-$0.58 & 0.18 & 4 & 5.43 \\
Fe~\textsc{i}   &   +4.33 & $-$3.17  &        0.00 &  0.09 & 99        & 7.50 \\
Fe~\textsc{ii}  &   +4.33 &  $-$3.17 &         0.00 &  0.07 & 12        & 7.50\\
Co~\textsc{i}  &+2.03 & $-$2.96 & +0.21 & 0.20 & 3 & 4.99 \\
Ni~\textsc{i}  &+3.01 & $-$3.21 & $-$0.04 & 0.03 & 3 & 6.22 \\
Zn~\textsc{i}  &+1.76 & $-$2.80  & +0.37 & 0.11 & 2 & 4.56 \\
Sr~\textsc{i}  &+0.05 & $-$2.82 & +0.35 & 0.10 & 1 & 2.87 \\
Sr~\textsc{ii} & $-$0.05 & $-$2.92 & +0.25 & 0.07 & 2 & 2.87 \\
Y~\textsc{ii} & $-$0.91 & $-$3.12 & +0.05 & 0.07 & 6 & 2.21 \\
Zr~\textsc{ii} & $-$0.21 & $-$2.79 & +0.38 & 0.06 & 2 & 2.58 \\
Ba~\textsc{ii} & $-$0.74 & $-$2.92 & +0.25 & 0.21 & 4  & 2.18 \\
La~\textsc{ii} & $-$1.52 & $-$2.62 & +0.55 & 0.13 & 6 & 1.10 \\
Ce~\textsc{ii} & $-$1.24 & $-$2.82 & +0.35 & 0.10 & 1 & 1.58 \\
Pr~\textsc{ii} & $-$1.67 & $-$2.39 & +0.78 & 0.10 & 1 & 0.72 \\
Nd~\textsc{ii} & $-$1.11 & $-$2.53 & +0.64 & 0.11 & 8 & 1.42 \\
Sm~\textsc{ii} & $-$1.30 & $-$1.99 & +1.18 & 0.06 & 7 & 0.69 \\
Eu~\textsc{ii} & $-$1.71 & $-$2.23 & +0.94 & 0.12 & 4 & 0.52 \\
Gd~\textsc{ii} & $<-$1.00 & $<-$2.07 &$<$+1.10 & \nodata& 1 & 1.07 \\
Tb~\textsc{ii} & $-$1.74 & $-$2.04 & +1.13 & 0.11 & 2 & 0.30 \\
Dy~\textsc{ii} & $-$1.00 & $-$2.10 & +1.07 & 0.09 & 2 & 1.10 \\
Er~\textsc{ii} & $-$1.50 & $-$2.42 & +0.75 & 0.10 & 1 & 0.92 \\
Hf~\textsc{ii} & $<-$1.22 & $<-$2.07 &$<$+1.10 &\nodata& 1 & 0.85 \\
Th~\textsc{ii} & $<-$2.05 & $<-$2.07 &$<$+1.10 &\nodata& 1 & 0.02 \\
\enddata      
\tablenotetext{a}{Solar photospheric abundances from Asplund et al. (2009).}
\end{deluxetable}

\begin{figure*}[t!]
\centering
\includegraphics[scale=0.56, trim = 11cm 0.5cm 11cm 0.6cm]{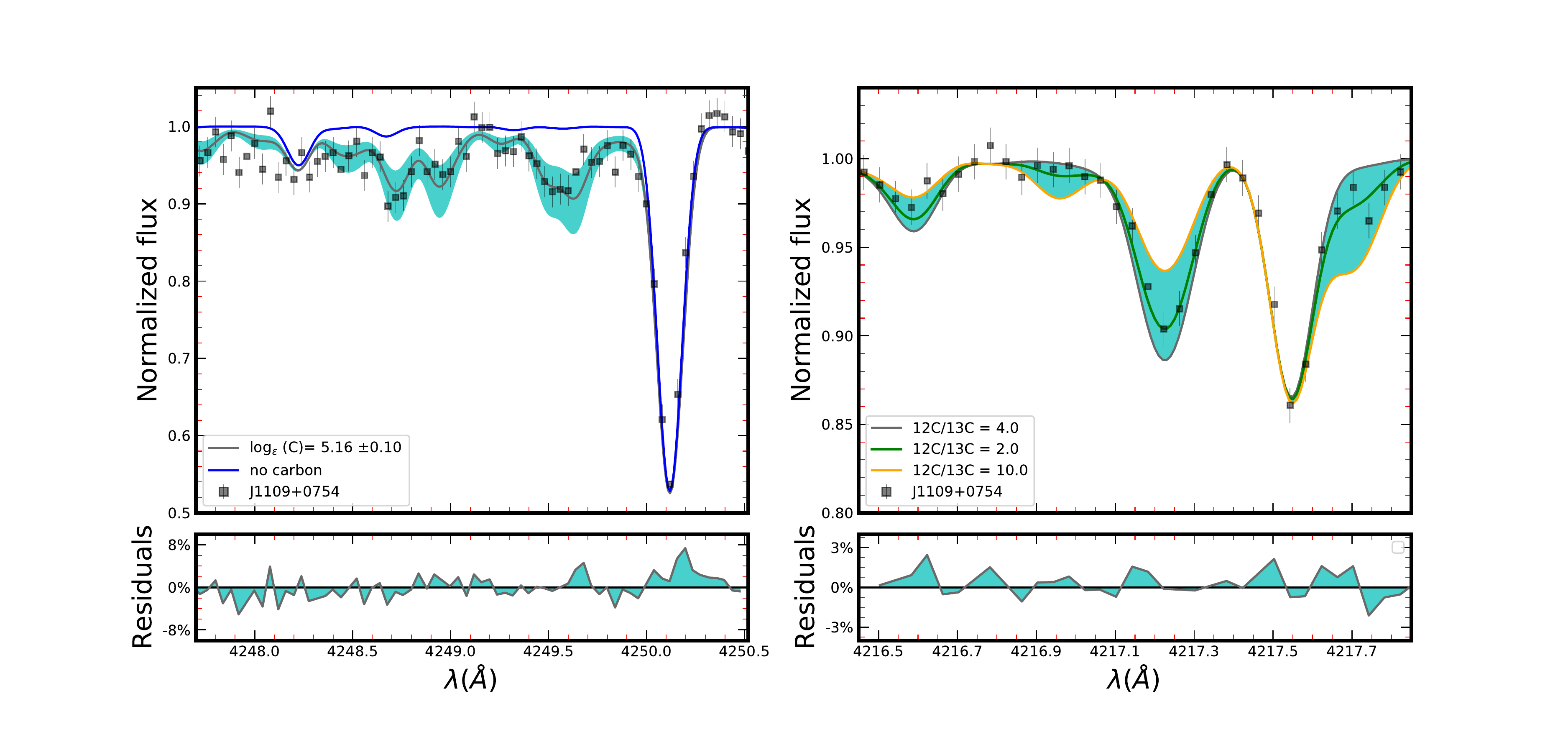}
\caption{Left panel: Portion of the J1109+0754 spectrum near the CH $G$-band. The filled squares indicate the observed data, and the solid red line denotes the best-fit carbon abundance. The turquoise shaded region encloses a $\pm 0.10$ dex difference in $\log \epsilon$ (C). Right panel: Determination of the carbon isotopic ratio $^{12}$C/$^{13}$C. The filled squares indicate the observed data, and the solid red line is the best-fit isotopic ratio. The lower panels represent the residuals between the best fits and the observed data.}
\label{fig:spectral1}
\end{figure*}

\section{Chemical Abundances}\label{sec:abundances}

We employed both equivalent-width analysis\footnote{Abundances were derived using MOOG's abfind driver} and spectral synthesis applied to the APF spectrum to derive the chemical abundances (and upper-limits) for a total of 31 elements, including 16 neutron-capture elements. The linelists for spectrum synthesis were generated with the linemake code\footnote{https://github.com/vmplacco/linemake}. The final adopted abundances are listed in Table~\ref{tab:abundances}. We use the Solar abundances from \citet{2009ARAA..47..481A} to calculate [X/H] and [X/Fe] ratios (where X denotes different elements). The standard deviation of the mean ($\sigma$), number of the lines used ($N_{\rm lines}$), and Solar abundances ($\log\varepsilon_{\odot}$) are also listed in Table~\ref{tab:abundances}. In the following, we comment on measurement details of individual elements and groups of elements. In Section~\ref{sec:discussion}, we further discuss the results and the abundance trends.

\subsection{Lithium}

J1109+0754 is on the upper red-giant branch, suggesting that some elements present in the stellar atmosphere have been altered due to the first dredge-up and other mixing processes. Lithium is affected by these processes, and its abundance is expected to be depleted compared to the Spite plateau value \citep[e.g.,][]{2016ApJ...819..135K}. This results in a weakened lithium doublet line at 6707\,{\AA}, making the derivation of an accurate abundances a challenge. Hence, we were only able to derive an upper limit using spectral synthesis. Our low upper limit of $A$(Li) $<-0.02$ agrees well with lithium values of other cool giants with similar temperatures \citep{2014AJ....147..136R}.

\subsection{Carbon and Nitrogen} 

Shortly after stars leave the main sequence and begin to ascend the giant branch, nucleosynthesis reactions in the core (e.g., the CN cycle) are expected to modify a number of the surface-element abundances (e.g., carbon, nitrogen, and oxygen). The most basic interpretation of the observed carbon and nitrogen, in evolved stars, is that their outer convective envelope expands and penetrates the CN-cycled interior through convective flows, which leads to an increase in the observed surface nitrogen and depletion in the surface carbon abundances \citep[e.g.,][]{1978ApJ...223L.117H, 1979A&A....78..323G, 1995ApJ...453L..41C, 2000A&A...358..671G, 2006A&A...455..291S, 2014ApJ...797...21P}. \textbf{(Added: However, when stars evolve past the luminosity bump, non-convective mixing can still be considered a viable source for dilution \citep{1967ZA.....67..420T,1968ApJ...154..581I})}.

We determined carbon abundances and isotope ratios ($^{12}$C/$^{13}$C) from spectrum synthesis by matching two portions of the CH $G$-band at $4280$\,{\AA}, and $4217$\,{\AA} with synthetic spectra of varying abundances and isotope ratios.  
This yielded a best-fit carbon abundance $\log \epsilon$ (C) $= 5.16 \pm 0.10$ (\abund{C}{Fe} = $-0.10$) and an \textbf{(Added: isotopic ratio $^{12}$C/$^{13}$C = 2.0 $\pm 2.0$. This low $^{12}$C/$^{13}$C ratio suggests that a significant amount of $^{12}$C has been converted into $^{13}$C}). Figure~\ref{fig:spectral1} shows the line fits used to obtain the carbon abundance and the $^{12}$C/$^{13}$C isotopic ratios (upper panels), and the residuals between the observed spectrum and the adopted best fits (lower panels). We use upper and lower carbon abundance fits ($\pm 0.10$\,dex) to assess the abundance uncertainty on the isotope ratio.  

Although we expected that J1109+0754 would exhibit an enhancement in its nitrogen abundance, we could not reliably detect any CN features in the APF spectrum, and thus determine a meaningful upper limit. 

\begin{figure*}[t!]
\centering
\includegraphics[scale=0.56, trim = 11cm 0.5cm 11cm 0.6cm]{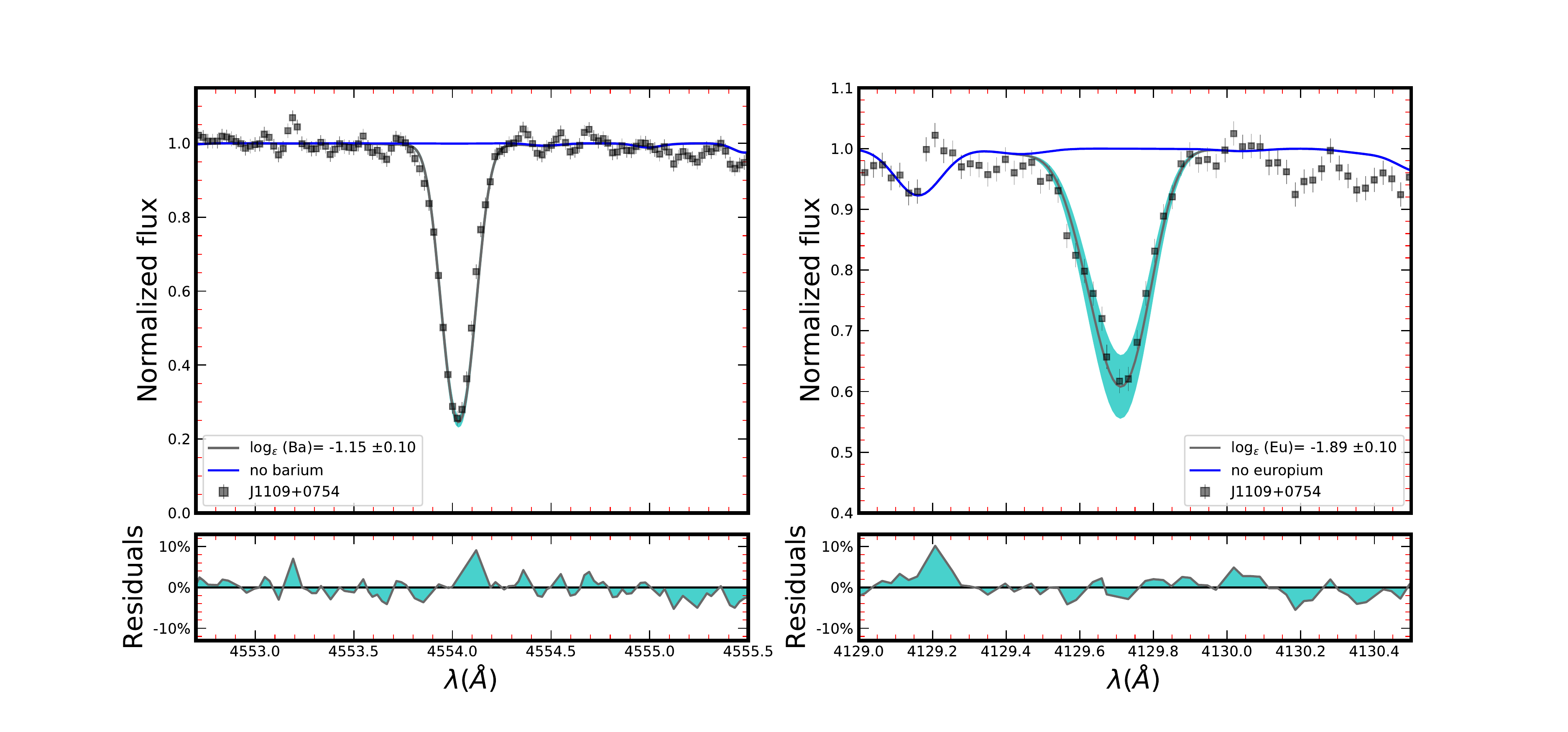}
\caption{Portions of the J1109+0754 spectrum near the \ion{Ba}{2} line at 4554\,{\AA} (left panel) and the \ion{Eu}{2} line at 4129\,{\AA}. Best-fit abundances obtained with spectrum synthesis are shown in the legends. Symbols and lines are the same as in Figure~\ref{fig:spectral1}.}
\label{fig:spectral2}
\end{figure*}

\subsection{Elements from Oxygen to Zinc}

We determined the atmospheric abundance of the light elements with different methods based on the availability of non-blended features with reliable continuum estimates. We only used equivalent-width analysis to derive the abundances of Ca, Sc, Ti, V, Cr, Mn, Co, Ni, and Zn. We used a combination of equivalent-width and spectrum-synthesis matching to measure the magnesium abundance. The O (6300 \,{\AA}), and Na (at 5889 \,{\AA} and 5895 \,{\AA}) were determined just from spectrum-synthesis. Table~\ref{tab:EWs} lists the atomic data, the measured EWs, and the atmospheric abundances for each individual line. Table~\ref{tab:abundances} lists the adopted average abundances.

\subsection{Neutron-Capture Elements}

We used spectrum synthesis to measure the abundances and upper limits for 16 neutron-capture elements (from Sr to Th). Where applicable, we took into account line broadening due to isotope shifts and hyperfine splitting structure\footnote{ Hyperfine structure information can be found in \\ https://github.com/vmplacco/linemake/blob/master/README.md}.

Figure~\ref{fig:spectral2} shows portions of the spectrum of J1109+0754 around the \ion{Ba}{2} line at 4554\,{\AA} and the \ion{Eu}{2} line at 4129\,{\AA}, and illustrates our technique of finding a best fit to the line. Spectrum matching the noise level of the line and continuum are used to determine the uncertainties.

We were able to measure the abundances for three elements that belong to the first \textit{r}-process peak, Sr, Y, and Zr. The abundances of strontium were measured using transitions from two different ionization stages (\ion{Sr}{1} at $4607.331$\,{\AA} and \ion{Sr}{2} at $4077.714$\,{\AA} and $4215.524$\,{\AA}); the results agree within 0.10\,dex. The abundances determined for yttrium, using six lines ($4235.731$\,{\AA}, $4358.727$\,{\AA}, $4883.684$\,{\AA}, $4900.110$\,{\AA}, $5087.420$\,{\AA}, and $5205.731$\,{\AA}), agree within 1-$\sigma$. For zirconium, we measured $\log \epsilon$ (Zr) = $-0.17$ and $-0.25$, using the two lines at 4149.198\,{\AA} and 4161.200\,{\AA}, respectively.

There are many absorption lines for neutron-capture elements within the spectral range of our data. However, many lines are located in blue regions with poor S/N, and some are heavily blended. Therefore, we used eight lines to measure $\log \epsilon$ (Nd) = $-1.11$, seven lines for $\log \epsilon$ (Sm) = $-1.30$, six lines for $\log \epsilon$ (La) = $-1.52$, four lines for $\log \epsilon$ (Ba) = $-0.74$ and $\log \epsilon$ (Eu) = $-1.71$, two lines for $\log \epsilon$ (Tb) = $-1.74$ and $\log \epsilon$ (Dy) = $-1.00$, one line for $\log \epsilon$ (Ce) = $-1.24$, $\log \epsilon$ (Pr) = $-1.67$, and $\log \epsilon$ (Er) = $-1.50$, and obtained upper limits for $\log \epsilon$ (Gd) = $< -1.0$,  $\log \epsilon$ (Hf) = $< -1.2$, and $\log \epsilon$ (Th) = $< -2.4$. Generally, line abundances for each element agree well with each other, which is reflected in the small reported standard deviations. Table~\ref{tab:abundances} lists our final abundances for all elements.

\subsection{Systematic Uncertainties}

\begin{deluxetable}{lrrrrrr} 
\tablenum{4}
\tablecaption{Systematic Abundance Uncertainties\label{tab:errors}}  
\tablewidth{0pt}
\tablecolumns{7} 
\tabletypesize{\scriptsize}
\tablehead{\colhead{Species}&\colhead{Ion}& \colhead{$\Delta$\mbox{T$_{\rm eff}$}}&\colhead{$\Delta\log g$}& \colhead{$\Delta v_{micr}$}&\colhead{Root Mean}\\ 
\colhead{}&\colhead{} &\colhead{+100\,K}& \colhead{$+$0.3\,dex}&\colhead{+0.3\,km\,s$^{-1}$}&\colhead{Square}} 
\startdata 
C &CH&   +0.19 & +0.11 &    0.00& 0.27 \\ 
O &1&    +0.22 & +0.11 &    0.00& 0.39 \\ 
Na& 1&   +0.09 &$-$0.01& $-$0.02& 0.10 \\ 
Mg& 1&   +0.09 &$-$0.06&   +0.06& 0.14 \\ 
Ca& 1&   +0.08 &$-$0.01& $-$0.03& 0.12 \\ 
Sc& 2&   +0.07 &  +0.11& $-$0.02& 0.15 \\ 
Ti& 1&   +0.10 &$-$0.01& $-$0.01& 0.14 \\ 
Ti& 2&   +0.04 &  +0.08& $-$0.12& 0.16 \\ 
V& 2&    +0.05 &  +0.10&   +0.00& 0.13 \\ 
Cr& 1&   +0.12 &$-$0.01& $-$0.03& 0.15 \\ 
Mn& 1&   +0.12 &  +0.00& $-$0.01& 0.13 \\ 
Fe& 1&   +0.12 &$-$0.02& $-$0.08& 0.16 \\ 
Fe& 2&   +0.02 &  +0.11& $-$0.01& 0.14 \\ 
Co& 1&   +0.12 &   0.00& $-$0.03& 0.15 \\ 
Ni& 1&   +0.15 &$-$0.03& $-$0.11& 0.20 \\ 
\hline
\hline
\tableline
\enddata
\end{deluxetable}

\begin{figure}
\epsscale{1.2}
\plotone{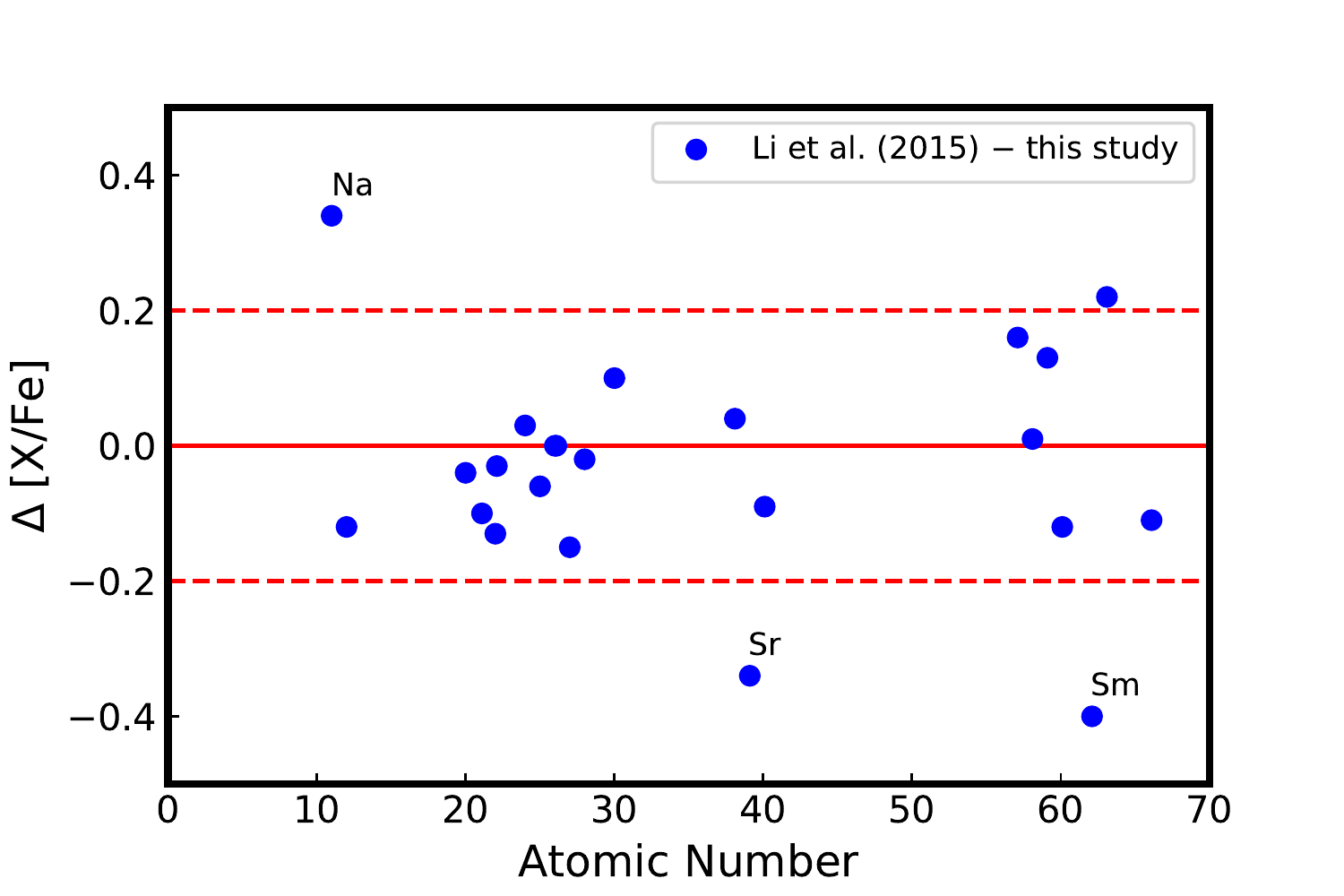}
\caption{Comparison of our abundances for the elements measured in J1109+0754 and in common with those presented in \citet{2015RAA....15.1264L}. The blue filled circles represent differences between abundances calculated as (\citealt{2015RAA....15.1264L} $-$ this work). The solid and dashed lines denote abundance differences of 0 and $\pm 0.20$\,dex, respectively.}
\label{fig:Haining}
\end{figure}

Table~\ref{tab:errors} lists the systematic uncertainties, as derived from varying the uncertainties of the adopted atmospheric models ($\Delta \teff = \pm150$K, $\Delta \logg = \pm 0.30$ dex, and $\Delta v_{\rm t} = \pm 0.30$\,km\,s$^{-1}$) one at the time. We then re-calculated our abundances. We take the differences as our final systematic uncertainties.

We also attempted to quantify systematic uncertainties associated with our measurement technique by comparing our results to those of \citet{2015RAA....15.1264L}. However, only a comparison of final derived abundances with those given in \citet{2015RAA....15.1264L} was possible. In Figure~\ref{fig:Haining}, we compare our \abund{X}{Fe} for species in common. Generally, there is good agreement of $-0.20 < $ $\Delta$ \abund{X}{Fe} $< +0.20$ dex. Still, sodium ($\Delta$ \abund{Na}{Fe} = +0.34 dex), strontium ($\Delta$ \abund{Sr}{Fe} = $-0.34$ dex), and samarium ($\Delta$ \abund{Sm}{Fe} = $-0.40$ dex) exhibited larger discrepancies. Taking into account differences in stellar parameters (\citealt{2015RAA....15.1264L} adopted \teff \,= 4440\,K, \logg\, = 0.70, \abund{Fe}{H} = $-$3.41, and $v_{\rm t}$= 1.98\,km\,s$^{-1}$) somewhat alleviates the discrepancies, but cannot fully reconcile them ($\Delta$ \abund{Na}{Fe} = +0.25 dex, $\Delta$ \abund{Sr}{Fe} = $-0.19$ dex, and $\Delta$ \abund{Sm}{Fe} = $-0.35$ dex), leaving potential differences in atomic data or measurement technique as a possible explanation.

\section{Discussion and Analysis}\label{sec:discussion}

\begin{figure*}[t!]
\centering
\includegraphics[scale=0.56, trim = 11cm 0.5cm 11cm 0.6cm]{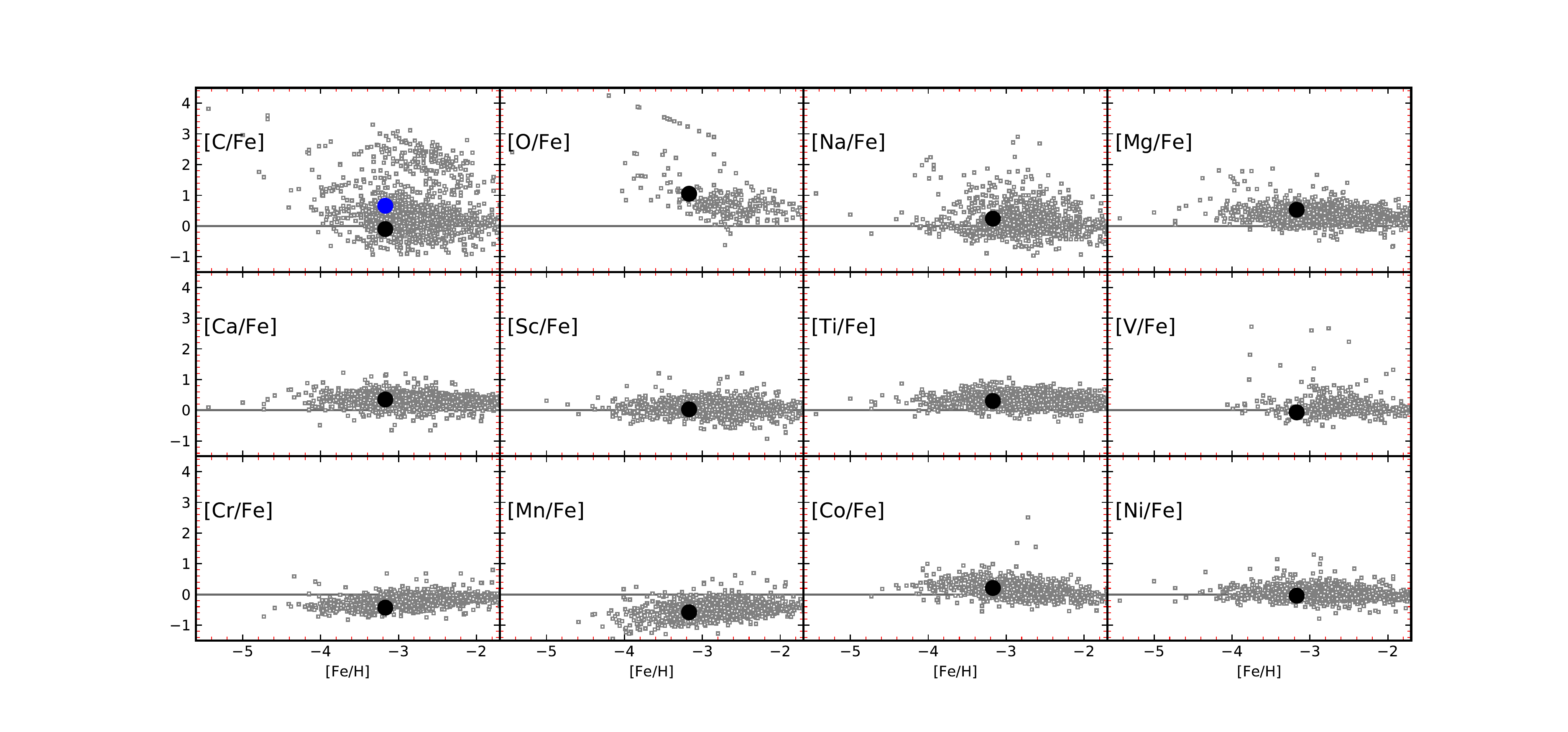}
\caption{Observed \abund{X}{Fe} ratios for elements up to the iron peak, as a function of \abund{Fe}{H}. Abundances for J1109+0754 are represented by the large filled black circles, those of Milky Way field stars with \abund{Fe}{H} $<-$2 as gray squares (JINAbase; \citealt{2018ApJS..238...36A}). The filled blue circle denotes the natal carbon abundance of J1109+0745, calculated based on \citet{2014ApJ...797...21P}.}
\label{fig:jina_light}
\end{figure*}

In this section, we evaluate the chemical-enrichment scenario for the natal gas cloud from which J1109+0745 formed. We were able to measure 31 individual elemental abundances (from lithium to thorium) for J1109+0754. Its neutron-capture elemental-abundance pattern indicates that this is an \textit{r}-II star, following the new definitions for RPE stars ([Eu/Fe] $> +0.7$ instead of [Eu/Fe] $> +1.0$ used previously) in \citet{2020arXiv200700749H}, based on new data collected by the RPA. Previously, it would have been considered as an $r$-I star, according to the definitions in \citealt{2005ARA&A..43..531B}. It is thus apparent that the overall abundance signature of J1109+0754 must have arisen after a variety of nucleosynthesis sources, including an $r$-process event, contributing to the elements we observe today.

\subsection{The Light-Element Abundance Pattern}

J1109+0754 has an approximately Solar \abund{C}{Fe} ratio. However, based on the \logg, it is assumed that J1109+0754 has undergone carbon depletion, and thus, the observed carbon abundance ($\abund{C}{Fe}$ = $-0.10$) does not reflect its natal value. 
Correcting for this significant effect \citep{2014ApJ...797...21P} increases the abundance to \abund{C}{Fe} = $+0.66$. Both the measured and corrected C abundances are shown in Figure~\ref{fig:jina_light}.
This suggests that J1109+0754 is close to the regime of strongly carbon-enhanced stars, and that significant amounts of carbon were present in its natal gas cloud.

Figure~\ref{fig:jina_light} shows the \abund{X}{Fe} abundances ratios of the light elements observed in J1109+0754. They all agree well with the abundances of Milky Way field stars taken from JINAbase \citep{2018ApJS..238...36A}. The observed $\alpha$-element (Mg, Ca, and Ti) abundances are enhanced with \abund{\alpha}{Fe} $\approx +0.4$, as it is typical for metal-poor halo stars.

Due to the extremely low-metallicity nature of J1109+0754 (\abund{Fe}{H} = $-$3.17 $\pm$ 0.09), we compare the observed light-element abundance pattern with the predicted nucleosynthetic yields of SNe for high-mass metal-free stars \citep{2010ApJ...724..341H}. This allows us to constrain the stellar mass and SN explosion energy of the progenitor of J1109+0754, assuming the gas was likely enriched by just one supernova \citep{Mardini_2019b}.

Using a normal distribution, we generated $10,000$ sets of the observed abundances up to the iron-peak from the corresponding measurement errors ($\sigma$, see Table \ref{tab:abundances}), resulting in  $10,000$ separate abundance patterns. We then used the online \texttt{STARFIT} code to find the best fit for each generated abundance pattern. These theoretical models have wide stellar-mass ranges (10-100 \ms), explosion energies ($0.3$-$10 \times 10^{51}$ erg), and $f_{mix}$ (no mixing to approximately total mixing)\footnote{http://starfit.org}. 

\begin{figure*}[t!]
\centering
\includegraphics[scale=0.70, trim = 11cm 0.5cm 11cm 0.1cm]{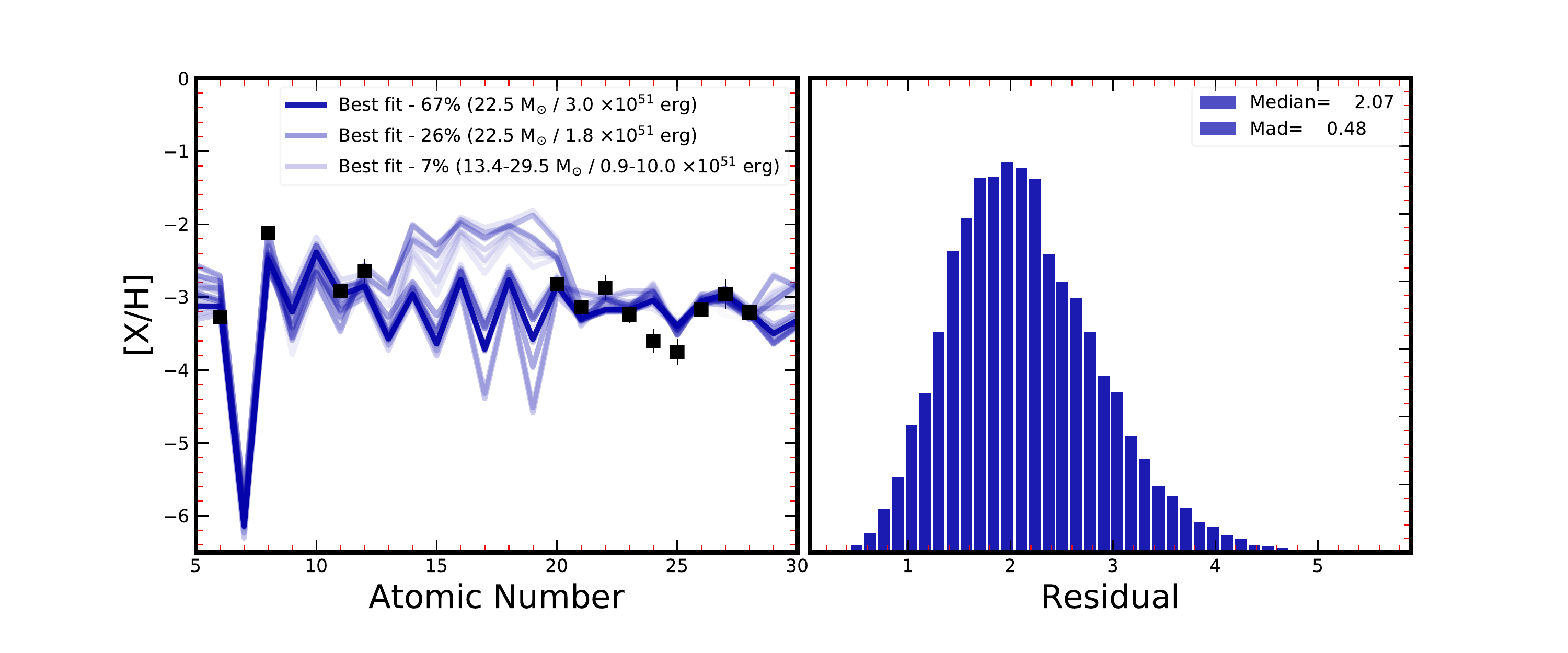}
\caption{(Left panel) The observed \abund{X}{H} abundance ratios of J1109+0754 (filled black squares), as a function of atomic number, overlaid with the matched predicted nucleosynthetic SNe models. The transparency of the models' lines reflect their fractional appearance. The best fits and their properties are discussed in the text. (Right panel) Posterior distributions for the mean squared residual, $\chi^{2}$, of the 10,000 simulations. The median and median absolute deviation (MAD) are shown in the legend.}
\label{fig:Starfit}
\end{figure*}

Figure~\ref{fig:Starfit} shows the best fits and their associated information. We found that $\approx 93\%$ of the generated patterns match the yields of two models with $22.5$\ms\, and explosion energies of 1.8 and 3 $\times 10^{51}$ erg. The remainder of the patterns ($\sim 7\%$) match 30 models with stellar masses ranging from $13.4$ to $29.5$ \ms and explosion energies ranging from $0.9$ to $10 \times 10^{51}$ erg. This agrees with results of \citet{Mardini_2019b}, and suggests that a single SN ejecta from a Population III stars with $22.5$\ms\ can be the responsible for the observed light elements pattern of J1109+0754. 

It thus appears that J1109+0754 may have formed in a halo that experienced the enrichment by a massive Population\,III star with a moderate explosion energy.

\subsection{The Heavy-Element Abundance Pattern}

\begin{figure*}[t!]
\centering
\includegraphics[scale=0.55, trim = 11cm 0.5cm 11cm 0.1cm]{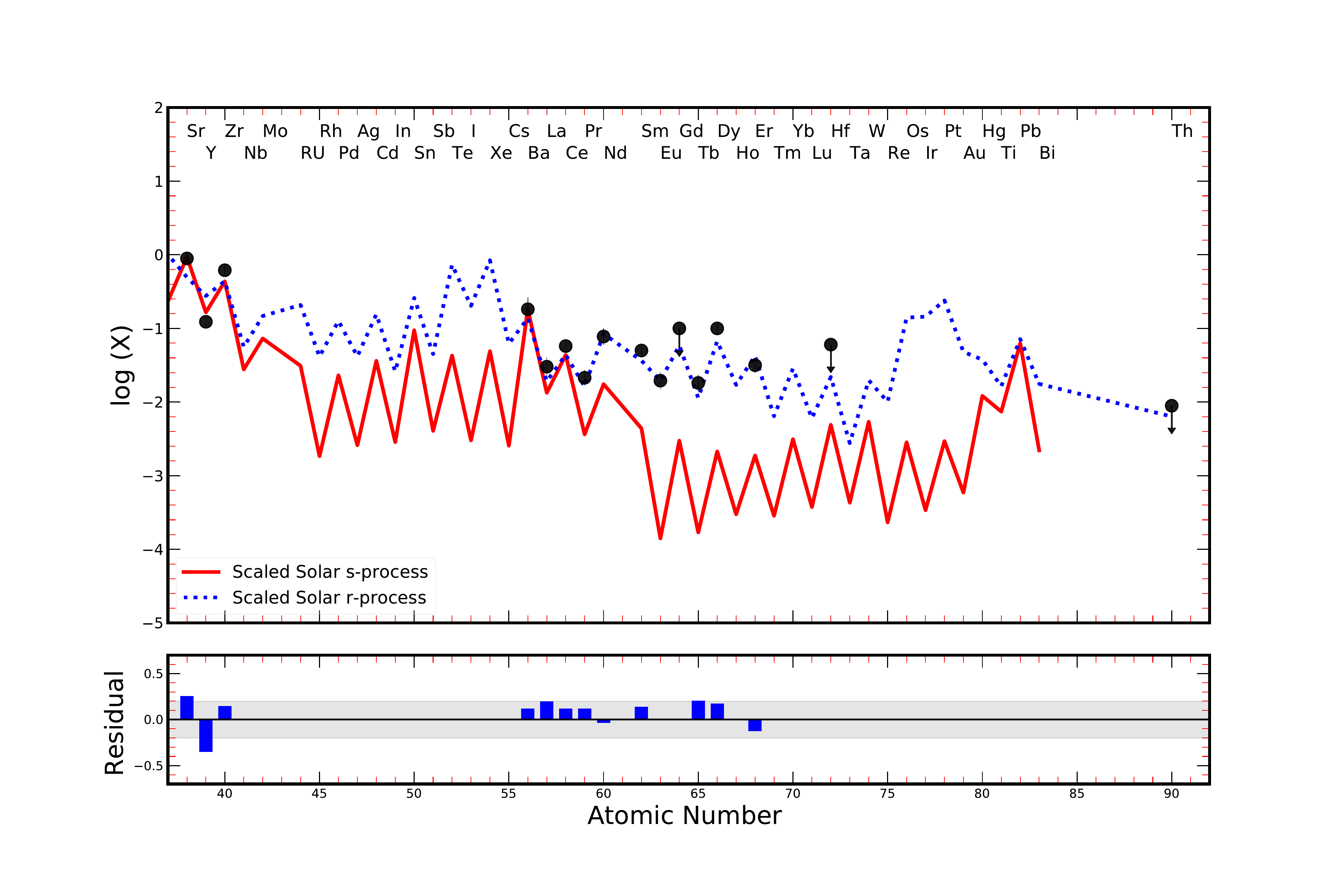}
\caption{Top panel: Heavy-element abundance pattern of J1109+0754 (filled circles), overlaid with the scaled Solar System abundances (SSSA). The Solar \textit{r}- and \textit{s}-process components (blue and red lines, respectively) are scaled to the observed Eu and Ba, respectively. The \textit{r}- and \textit{s}-process fractions are adopted from \citep{Burris_2000}. Lower panel: Residuals between the observed and the SSSA patterns.}
\label{fig:heavy_pattern}
\end{figure*}

The heavy-element abundances provide key information on the nucleosynthetic sources that operated in the birth environments of metal-poor stars (e.g., formation rates, timescales). Figure~\ref{fig:heavy_pattern} (top panel) shows the full \textit{r}-process abundance pattern of J1109+0754, overlaid with the scaled Solar System \textit{r}- and \textit{s}-process components (scaled to Eu and Ba, respectively). The \textit{r}- and \textit{s}-process fractions are adopted from \citet{Burris_2000} and isotopic ratios from \citet{2008ARA&A..46..241S}. 

The observed heavy-element abundances clearly match the pattern of the main $r$-process, as evidenced by the small standard deviation of the residuals (observed  abundances $-$ scaled-Solar \textit{r}-process pattern). This result supports the universality of the main \textit{r}-process, as has already been seen for many other metal-poor RPE stars, independent of their metallicity \citep[e.g.,][]{2002A&A...387..560H,2003ApJ...591..936S,2007ApJ...660L.117F,2018ApJ...865..129R,2020arXiv200604538P}. With $\abund{Eu}{Fe}= +0.96$, J1109+0754 is thus confirmed as an $r$-II star. In fact, given its  enhanced natal carbon abundance (\abund{C}{Fe} = +0.66 $\pm 0.27$), J1109+0754 adds to the sample of known \cemp \textit{r}-II star.

Deriving abundances of the actinide elements (thorium and uranium) would enable nucleo-chronometric age estimates of J1109+0754. However, only an upper limit for the thorium abundance could be determined, and no uranium features were detected, thus precluding any age measurements.

Finally, we note for completeness that the derived abundances of Sr, Y, and Zr do not match the scaled-Solar \textit{r}-process pattern as well. Similar variations have been observed in other RPE stars. The scatter in these abundances likely stems from differences in the yields produced by the limited, or weak, \textit{r}-process \citep{1996ApJ...467..819S, 2008ARA&A..46..241S, 2018ARNPS..68..237F}, or other $r$-process components.

\section{Kinematic Signature and Orbital Properties of J1109+0754}\label{sec:kinematics}

\begin{figure*}[t!] 
\centering
\makebox[\textwidth]{\includegraphics[width=\paperwidth]{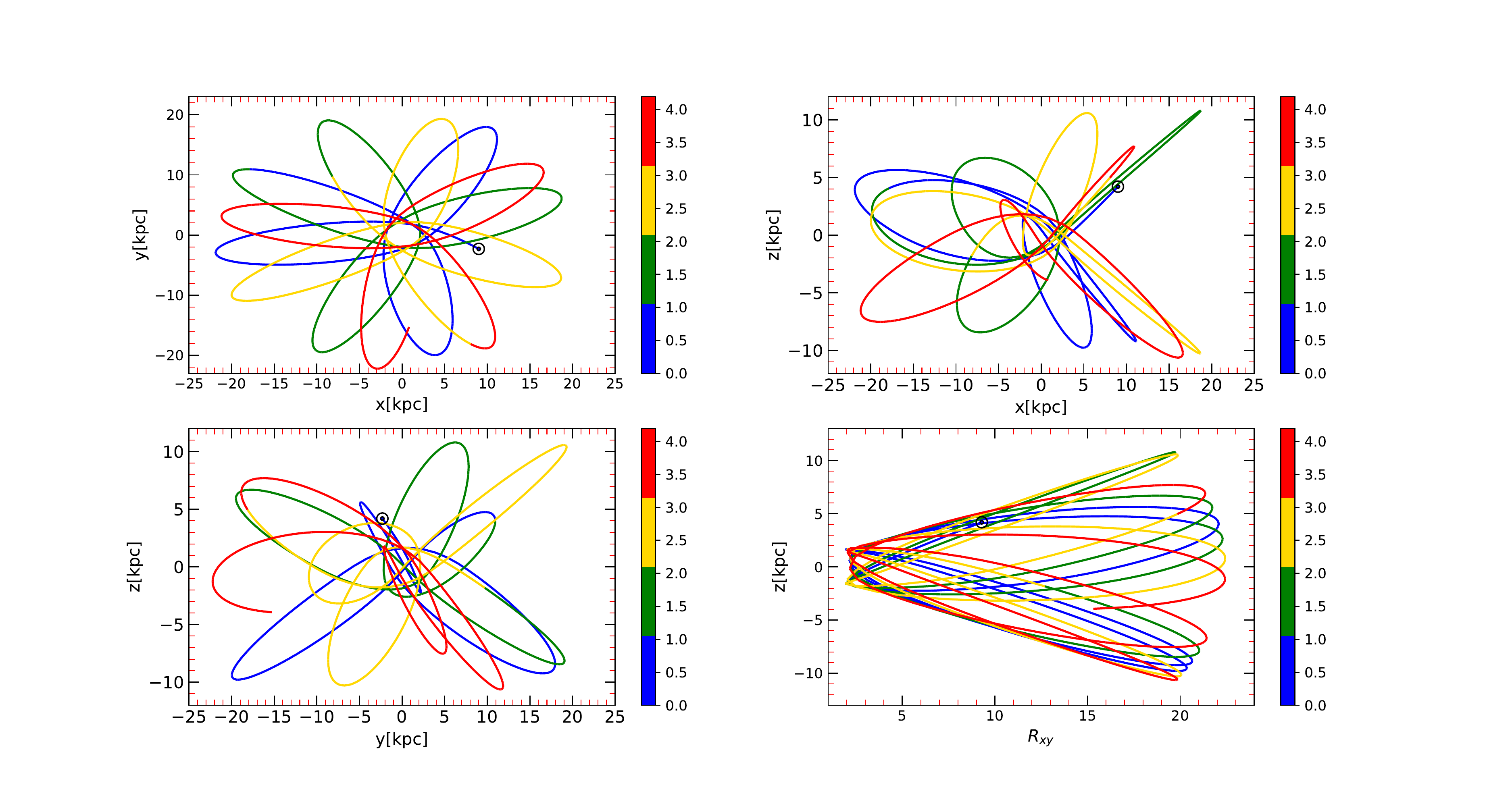}}
\caption{Orbits of J1109+0754, integrated in time for 4\,Gyr in the Galactic potential \texttt{MWPotential2014}. The upper panel shows $Y_{GC}$ (left) and $Z_{GC}$ (right), as a function of $X_{GC}$. The lower panel show $Z_{GC}$ (left ), as a function of $Y_{GC}$ and Galactocentric radius, $R_{\rm xy}$. The black points  in all panels indicate the current location of J1109+0754.}
\label{fig:Orbits-static}
\end{figure*}

The advent of the $Gaia$ mission \citep[Gaia DR2;][]{2018A&A...616A...1G} has fundamentally changed our view of the nature of the Milky Way. Detailed orbits for many stars can now be obtained from astrometric solutions provided by DR2, which have since led to the discovery of important structures in our Galaxy, e.g., the $Gaia$-Enceladus-Sausage \citep{2018MNRAS.478..611B,2018ApJ...863..113H,2018Natur.563...85H, 2020arXiv200608625N} and the Sequoia event \citep{2019MNRAS.488.1235M}. This followed numerous hints over several decades for the presence of such structures, primarily provided by results from spectroscopy, photometry, and, in some cases, proper motions, using smaller samples of metal-poor stars (e.g, \citealt{1986ApJS...61..667N, 1997ApJ...481..775S, 2000AJ....119.2843C}). The $Gaia$ results identified the {\it reasons} for these complexities, validated the previous claims, and gave names to several prominent examples. Other studies \citep[e.g.,][]{2007Natur.450.1020C,2010ApJ...712..692C,2012ApJ...746...34B,2020ApJ...897...39A} then presented evidence for the halo being built by multiple components involving at least an inner- and an outer-halo population.

To investigate the kinematic signature and orbital properties of J1109+0754, we adopted the parallax and proper motion from \citet{2018A&A...616A...1G}, the distance from \citealt{2018AJ....156...58B}, and the radial velocity derived from the APF spectrum as the line-of-sight velocity (see Table~\ref{tab:observations}). \textbf{(Added: We have taken into account the known zero-point offset parallax for bright stars in $Gaia$ DR2, as described in \citet{2018A&A...616A...2L})}.

\subsection{Orbital Properties with \texttt{AGAMA}}\label{sec:AGAMA}

Unraveling the full kinematic signature of J1109+0754 enables us to learn about its formation history, and potentially the environment in which the star formed. Using the public code \texttt{AGAMA} \citep{2019MNRAS.482.1525V} with the fixed Galactic potential \texttt{MWPotential2014} \citep[see][for more information]{2015ApJS..216...29B} we thus integrate the detailed orbital parameters available for J1109+0754. For that, we generated 10,000 sets of the six-dimensional phase space coordinates based on the corresponding measurement uncertainties ($\sigma$, see Table \ref{tab:observations}), to then statistically derive the total orbital energy, and calculate the three-dimensional action ($\vector{J}=(J_r, J_\phi, J_z)$). We also calculate Galactocentric Cartesian coordinates ($X_{GC}, Y_{GC}, Z_{GC}$), Galactic space-velocity components $(U, V, W)$, and cylindrical velocities components (V$_{R}$, V$_{\phi}$, V$_{z}$) \textbf{(Added: are defined in the same way as presented in \citet{Mardini_2019b, Mardini_2019a})}.

For our calculations, we assume that the Sun is located on the Galactic midplane ($Z_{\odot}= 0$) at a distance of $R_{\odot}=8.0$\,kpc from the Galactic center \citep{2010ASPC..438...16F}. The local standard of rest (LSR) velocity at the Solar position is $v_{LSR} =$~232.8\,\mbox{km\,s$^{\rm -1}$} \citep{2017MNRAS.465...76M}, and the motion of the Sun with respect to the LSR is $(U_\odot, V_\odot, W_\odot) =$ (11.1, 12.24, 7.25)\,\mbox{km\,s$^{\rm -1}$} \citep{10.1111/j.1365-2966.2010.16253.x}. We define the total orbital energy as $E = (1/2) \vector{v}^2 + \Phi(\vector{x})$, the eccentricity as $e = (\rapo - \rperi) / (\rapo + \rperi)$, the radial and vertical actions ($J_r$ and $J_z$, respectively) to be positive \citep[see][for more information]{10.1111/j.1365-2966.2012.21757.x}, and the azimuthal action by:

\vspace{0.3cm}
\begin{equation}
J_\phi = \frac{1}{2 \pi} \oint_{\rm orbit} \mathrm{d}\phi \; R V_\phi = - L_z
\end{equation}
\vspace{0.3cm}

Table~\ref{tab:UVW} lists the calculated median for the Galactic positions and Galactic velocities. Table~\ref{tab:kinematic} lists the medians for the orbital energy, orbital parameters, and the three-dimensional actions. The sub- and superscripts denote the 16th and 84th percentile confidence intervals, respectively, for each of these quantities. 

These results show that J1109+0754 possesses a bounded (E $< 0$), non-planar ($J_z \neq 0 $ and $\zmax \neq 0$), and eccentric ($J_r \neq 0 $ and $e \neq 0$) orbit. Moreover, the positive $J_\phi$ and $V_\phi$ values (see Table~\ref{tab:UVW}) indicate that J1109+0754 is on a prograde orbit. Figure~\ref{fig:Orbits-static} shows the last ten orbital periods of J1109+0754, in different projections (XY, XZ, and YZ) onto the Galactic plane, integrated for 4.2\,Gyr. The minimum distance of J1109+0754 from the center (\textbf{(Added: pericenter = $1.9^{0.9}_{0.8}$\,kpc), the maximum distance (apocenter = $21.7^{2.9}_{2.3}$\,kpc), and the maximum height of the star above the Galactic plane ($\zmax = 10.9^{1.8}_{1.8}$\,kpc))} suggest that the orbit of our star reaches out to distances of the inner-/outer-halo overlap region.

\begin{deluxetable*}{lDDDcDDDcDDDrr}
\tablenum{5}
\tablecaption{Positions and Galactic Space-Velocity Components \label{tab:UVW}}
\tablewidth{0pt}
\tabletypesize{\scriptsize}
\tablehead{
\colhead{Star} &\twocolhead{X} &\twocolhead{Y} &\twocolhead{Z} &\colhead{} &\twocolhead{U} &\twocolhead{V} &\twocolhead{W} 
&\colhead{} &\twocolhead{$V_{R}$} &\twocolhead{$V_{\phi}$}&&\twocolhead{$V_{\perp}$}  \\
\cline{2-7}\cline{9-14}\cline{16-23} \colhead{} &\multicolumn{6}{c}{(kpc)} &\colhead{} &\multicolumn{6}{c}{(km s$^{-1}$)} &\colhead{} 
& &\multicolumn{6}{c}{(km s$^{-1}$)} &\colhead{}  }
\decimals
\startdata
J1109+0754 & 8.97^{+0.14}_{+0.14}       &      $-$2.32^{+0.32}_{+0.32}       &4.22^{+0.28}_{+0.28}       & &$-$170.12^{+9.3}_{+9.4}   &119.92^{+13.81}_{+13.13}   &$-$133.45^{+7.8}_{+7.6}     & &$-$194.64^{+16.29}_{+14.55}   & 73.72^{+7.82}_{+8.01}   &  235.99^{+4.31}_{+4.39}    \\
\hline
\hline
\tableline
\enddata
\tablecomments{The $-$ and $+$ indicate the 16th percentile and 84th percentiles}
\end{deluxetable*}
 \begin{deluxetable*}{lDDDcDDDcDDDrr}
\tablenum{6}
\tablecaption{Derived Dynamical Parameters \label{tab:kinematic}}
\tablewidth{0pt}
\tabletypesize{\scriptsize}
\tablehead{
\colhead{Star} &\twocolhead{$r_\ensuremath{\mathrm{peri}}$} &\twocolhead{$r_\ensuremath{\mathrm{apo}}$} &\twocolhead{$Z_\ensuremath{\mathrm{max}}$} &\colhead{} &\twocolhead{e} &\twocolhead{E}  &\colhead{} &\twocolhead{ L$_z$} 
& \twocolhead{J$_r$} &\twocolhead{J$_z$} &\twocolhead{J$_\phi$} \\
\colhead{} &\multicolumn{6}{c}{(kpc)} &\colhead{} &\multicolumn{6}{c}{($10^{3}$ km$^{2}$ s$^{-2}$)} &\colhead{} 
& &\multicolumn{6}{c}{(kpc km s$^{-1}$)} &\colhead{}  }
\decimals
\startdata
J1109+0754 & 1.88^{+0.89}_{+0.82}       &      21.66^{+2.90}_{+2.25}       &10.87^{+1.78}_{+1.83}       & &0.84^{+0.08}_{+0.09}   &$-$87.05^{+4.12}_{+3.26}   &683.68^{+72.95}_{+89.74}     & &1226.86^{+178.21}_{+113.70}   & 224.04^{+5.34}_{+5.89}   &  683.68^{+72.95}_{+89.74}  &\\
\hline
\hline
\tableline
\enddata
\tablecomments{The $-$ and $+$ indicate the 16th percentile and 84th percentiles}
\end{deluxetable*}

\subsection{Detailed Kinematic History of J1109+0754}\label{sec:orbit_integr}

The kinematic results from Section~\ref{sec:AGAMA} are insightful, but limited, given that fixed Galactic potential has been used. Below we explore a novel approach with the goal to obtain a less-idealized and more-realistic time-dependent kinematic history of J1109+0754 within the Milky Way halo, which itself was built from smaller accreted dwarf galaxies, one of which likely contributed this star to the halo at early times.

In order to do so, we combine our custom high-order Hermite4 code \PGRAPE \citep{HGM2007}\footnote{The current version of the \PGRAPE code is available here {\tt ftp://ftp.mao.kiev.ua/pub/berczik/phi-GRAPE/}}. It
uses a GPU/CUDA-based GRAPE emulation YEBISU library \citep{NM2008}. It has been well-tested and used with several large-scale (up to few million particles) simulations \citep{2016MNRAS.460..240K, 2014ApJ...780..164W, 2014ApJ...792..137Z, 2012ApJ...758...51J, 2012ApJ...748...65L,  2020arXiv200601960M}. \textbf{(Added: We selected the particle data of Milky Way analogs from the publicly available \texttt{Illustris-TNG} cosmological simulation set)} \citep{2015MNRAS.449...49R,2018MNRAS.480.5113M, 2018MNRAS.477.1206N,  2018MNRAS.475..624N, 2018MNRAS.475..648P, 2018MNRAS.475..676S, 2019MNRAS.490.3234N, 2019ComAC...6....2N, 2019MNRAS.490.3196P}. Our goal is to establish a {\it time-varying} potential, based on the subhalo's snapshot data for all redshifts, and then carry out a detailed integration of J1109+0754's orbital parameters over some 10\,Gyr backward in time.

\subsubsection{Selecting Milky Way-like Galaxies in \texttt{Illustris-TNG}}

We use the \texttt{Illustris-TNG} TNG100 simulation box, characterized by a length of $\sim 110$ $\mathrm{Mpc}$. TNG100 is the second highest resolution simulation box among the TNG simulations, and has the highest resolution of the publicly available data. The TNG100 simulation box is thus sufficiently large to contain many resolved Milky Way-like disk galaxies. The mass resolution in TNG100 is $7.5 \times 10^6\,\rm{M}_{\odot}$ and  $1.4 \times 10^6\,\rm{M}_{\odot}$ for dark matter and baryonic particles, respectively. This resolution is larger, by a factor $\sim 10$, than the corresponding mass resolution of particles in TNG300, allowing a more accurate description of structure formation and evolution. Given that Milky Way-like galaxies have dark matter halos of $\sim 10^{12}\,\rm{M}_{\odot}$ and disks with $\sim 10^{10}\,\rm{M}_{\odot}$, we identify simulated galaxy candidates, with $10^5$ to $10^6$ dark matter particles and $10^3$ to $10^4$ stellar particles, to ensure good particle-number statistics.

To select potential Milky Way analogs, we target all $z=0$ galaxies (subhalos) which reside in the centers of massive halos with $0.6\times 10^{12} <M_{200}/{\rm M}_\odot<2\times 10^{12}$ (where $M_{200}$ is defined as the total halo mass in a sphere whose density is 200 times the critical density of the Universe; \citealt{1995MNRAS.275...56N}). This mass range reflects literature values \citep[e.g,][]{2017MNRAS.467..179G,2020MNRAS.491.3461B}, and corresponds to stellar masses in accordance with observations \citep{2011MNRAS.414.2446M}. 

We then pared down an initial list of over 2000 candidates using the following criteria: i) Stellar subhalo mass: We enforce a $3\sigma$ range of the Milky Way stellar mass \citep{2015JCAP...12..053C}, by only counting galaxies with a total stellar mass of $4.5\times 10^{10}{\rm M}_\odot < M < 8.3\times 10^{10} {\rm M}_\odot$. ii) Morphology: To account for the disky structure of the Milky Way, we select only galaxies with a triaxiality parameter $T< 0.35$, which we define as $T = \sqrt{(c/a)^2+(1-(b/a))^2 }$, where $a, b, c$ are the principal axes of inertia. iii) Kinematics: To select disky galaxies, we also calculate the circularity parameter of each stellar particle, $\epsilon = j_{z}/j$, where $j_z$ is the specific angular momentum along the z axis, and $j$ is the total specific angular momentum of the star \citep{2003ApJ...597...21A}, respectively. Only galaxies which have at least 40\% of stars with $\epsilon > 0.7$ \citep{2014MNRAS.437.1750M} are counted. iv) Disk-to-total mass ratio: We only select galaxies with stellar populations with disk-to-total mass ratios between 0.7 and 1, corresponding to the Milky Way's ratio of 0.86 \citep{2011MNRAS.414.2446M}. Stars with $\epsilon > 0.6$ are assigned to the disk.

By applying the above criteria, we obtain a total of 123 Milky Way-like candidates. We emphasize that the goal of our study is not to test the capability of TNG100 to reproduce Milky Way-like galaxies. Therefore, we acknowledge that these criteria, while sufficient for our work, do not necessarily reflect the true number of Milky Way analogs in TNG100. Unfortunately, not all of these subhalos are equally useful for establishing a time-dependent potential: If parameters change too much between consecutive snapshots, the resulting interpolation is not accurate enough. This typically occurs as the result of the subhalo switching problem \citep[see description in][]{2017MNRAS.472.3659P}, where a group of loosely bound particles is included in one of our subhalos in some snapshots (by the Subfind algorithm) but excluded in others. We finally chose subfive halos as the most suitable Milky-Way analogs for our study, and use their corresponding potentials for our orbital integrations.

\begin{figure*}
\centering
\includegraphics[width=0.95\columnwidth]{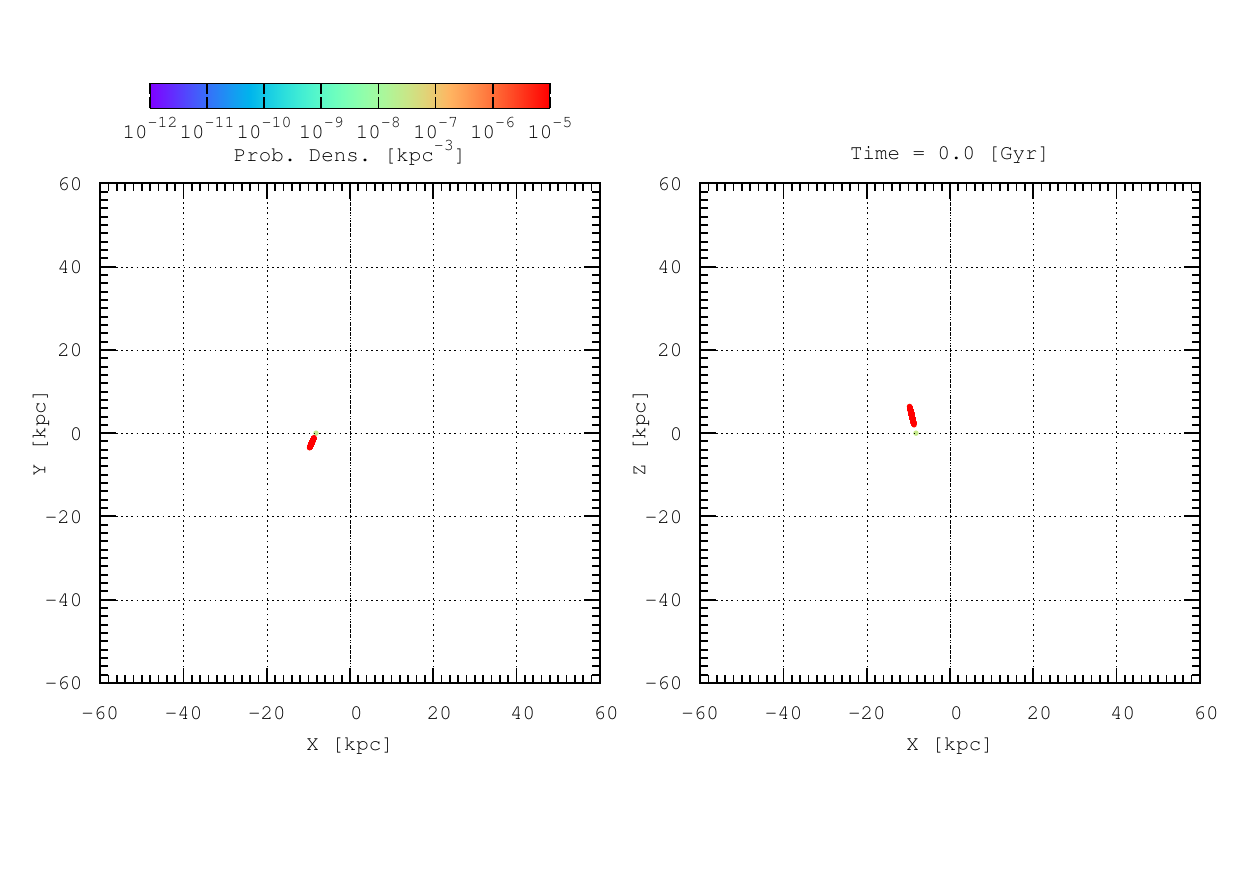}
\includegraphics[width=0.95\columnwidth]{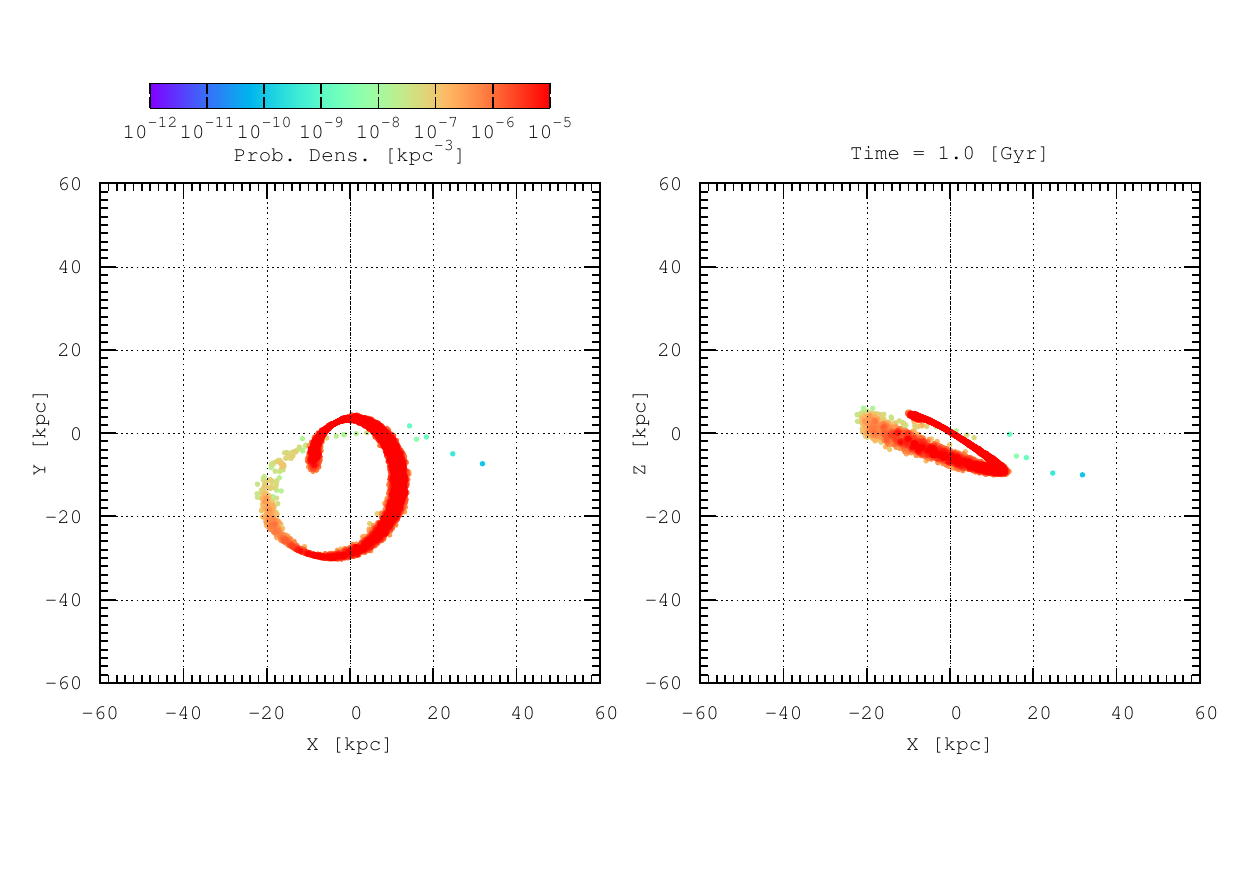}
\includegraphics[width=0.95\columnwidth]{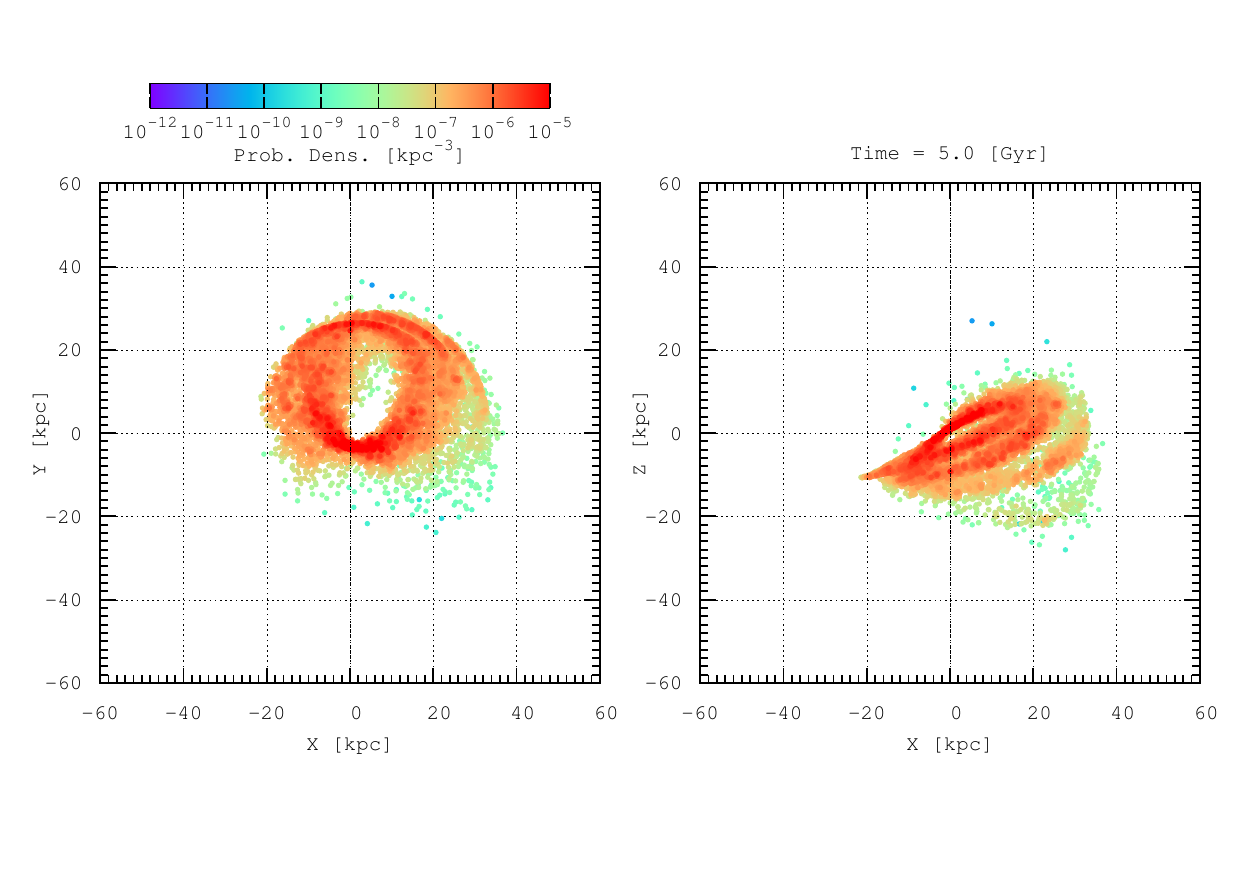}
\includegraphics[width=0.95\columnwidth]{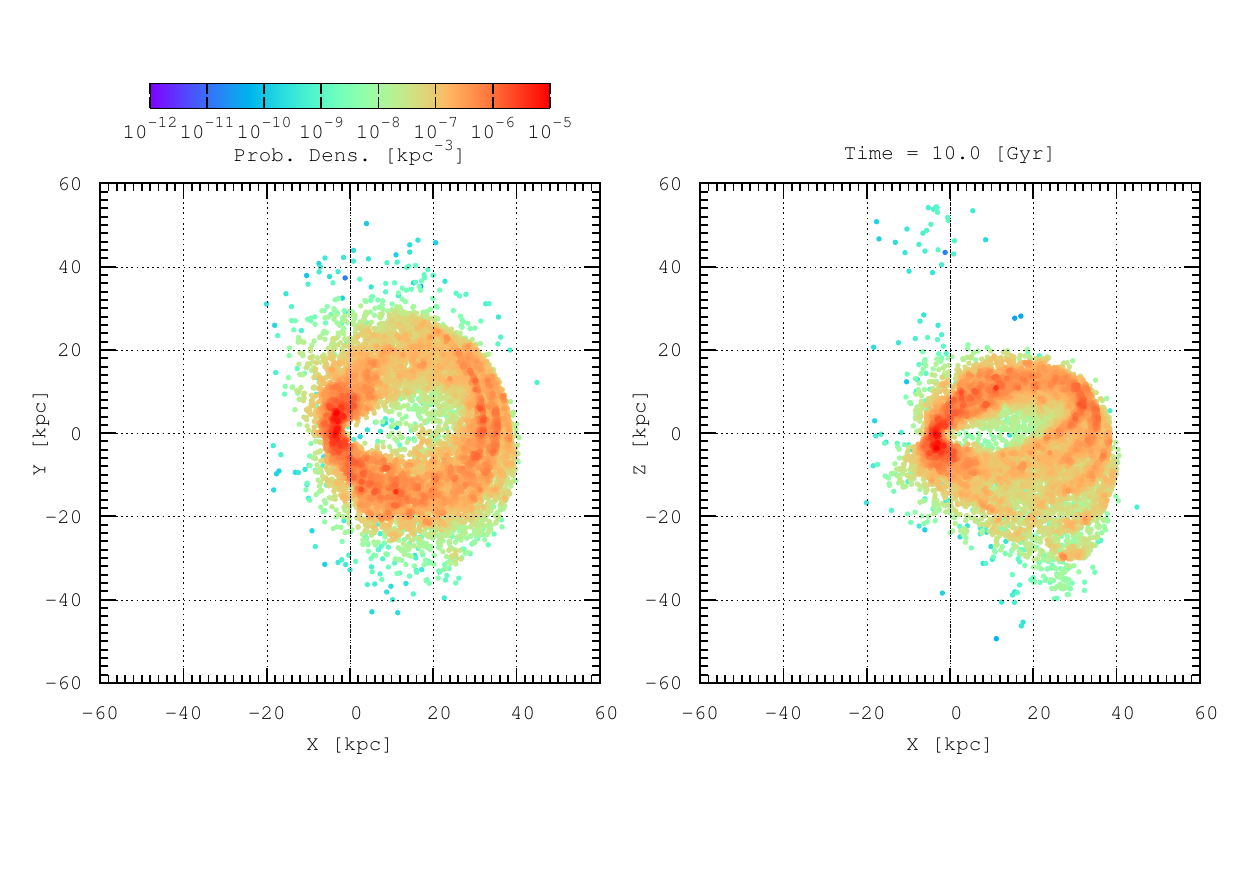}
\caption{Results of the backward integration (for 10\,Gyr) of LAMOST J1109+0754's orbital parameters following 10,000 random coordinates and velocity realizations (inside $\pm~2\sigma$) for  \texttt{Illustris-TNG} halo {\tt \#489100}. 
The color-coding captures the probability number density of the orbits in a 1\,kpc$^3$ cube at each point.}
\label{fig:orbits-10k}
\end{figure*}

\subsubsection{Establishing the ORIENT}

To obtain a time-varying potential for each Galactic analog, we model the gas and dark matter particles as a single Navarro–Frenk–White (NFW) \citep{1995MNRAS.275...56N} sphere, and the stellar particles as a \citet{1975PASJ...27..533M} disk. The best-fitting parameters are found for each snapshot using a smoothing spline fitting procedure (see the Appendix). The complete potential for each subhalo across all snapshots is then a set of parameters for the disk and the stellar halo as obtained from each TNG100 snapshot. To obtain the gravitational force as a function of time, we interpolate between each snapshot's parameters. Effectively, all the individual integrations for each snapshot are {\it not} one true $N$-body simulation, but a series of independent 1-body simulations or scattering experiments (i.e., test-particle integration in a external potential). To efficiently perform these experiments with 10,000 particles at once, we take advantage of the parallel framework of the \PGRAPE code.

To find the optimal integration parameter, $\eta$, which drives the integration accuracy of our code, we first ran short (up to 1\,Gyr) test simulations with different values of $\eta$ (0.020, 0.010, 0.005, 0.001) using a Milky Way fixed model, taken from \citet{2011A&A...536A..64E} (see also their Figure~1).
Using $\eta = 0.01$ limits the total relative energy drift ($\rm{dE}_{\rm tot}/\rm{E_{\rm tot}(t=0)}$) in a 10\,Gyr forward integration in the fixed Milky Way potential to below $\approx 2.5 \cdot 10^{-13}$, thus optimizing code speed vs. accuracy.

\subsubsection{Results}

Figure~\ref{fig:orbits-10k} shows the results of the backward orbital 
integrations of the 10,000 realizations in our selected \texttt{Illustris-TNG} subhalo {\tt \#489100}. It is clear that after $\sim$ 1\,Gyr, the initial positions of the random cloud extend over the entire model galaxy range. 
The color coding in the figure shows the probability number density of the orbits in a 1\,kpc$^3$ cube at each point. After 10\,Gyr of backward integration,  the positions of some realizations extend up to $\sim$ 60\,kpc from the center. This implies that they are already a part of the outer halo of our simulated galaxy. This finding supports the likely external origin (larger galaxy merger or smaller dwarf tidal disruption) of J1109+0754. 

\begin{figure*}
\centering
\includegraphics[width=0.65\columnwidth]{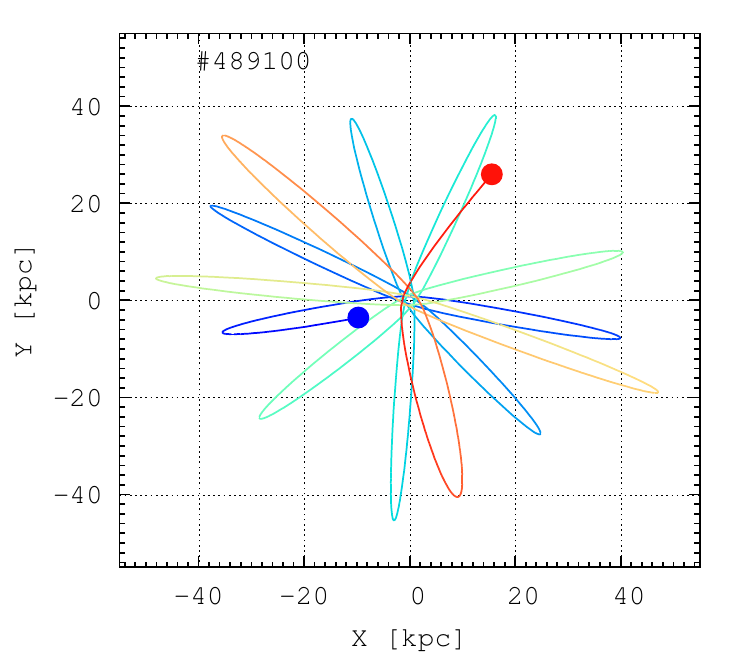}
\includegraphics[width=0.65\columnwidth]{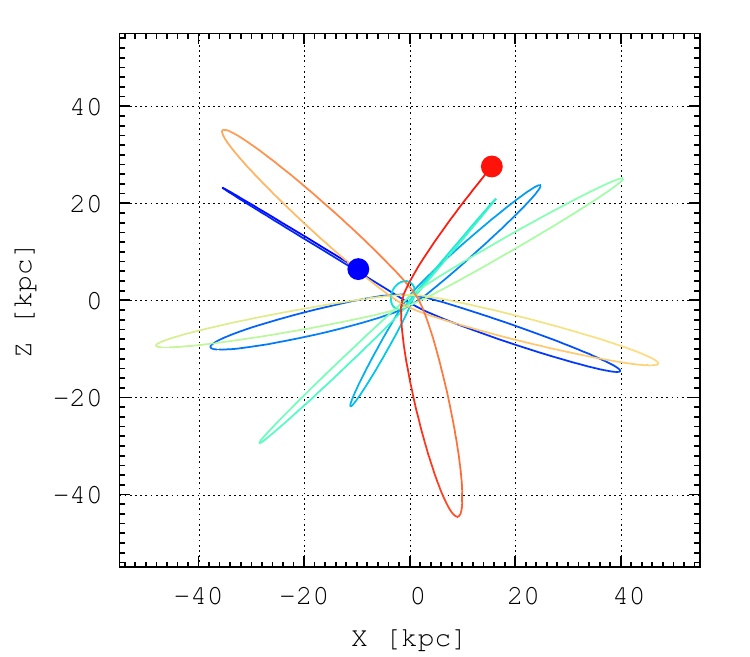}
\includegraphics[width=0.65\columnwidth]{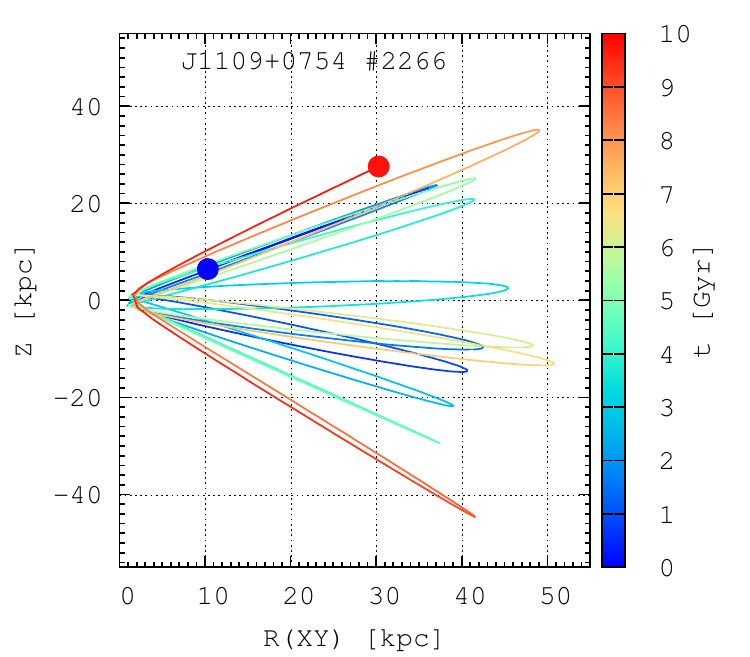}
\includegraphics[width=0.65\columnwidth]{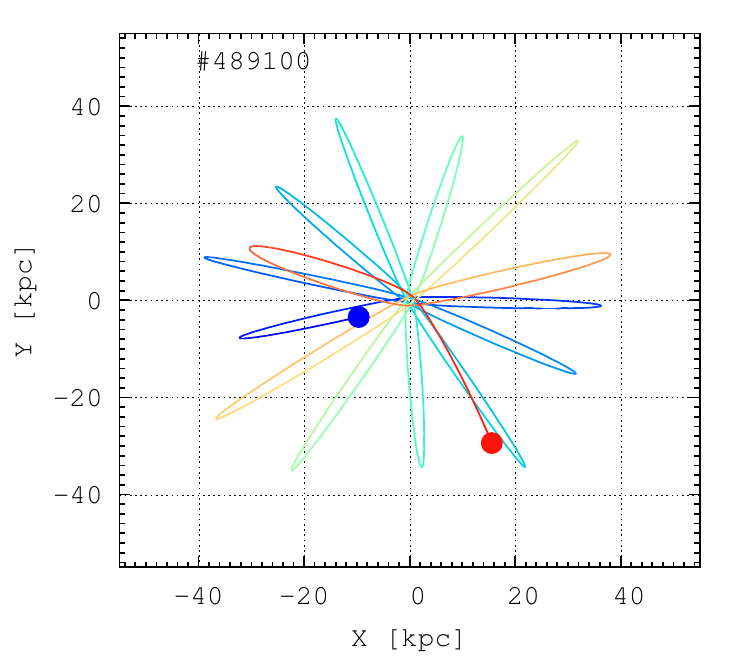}
\includegraphics[width=0.65\columnwidth]{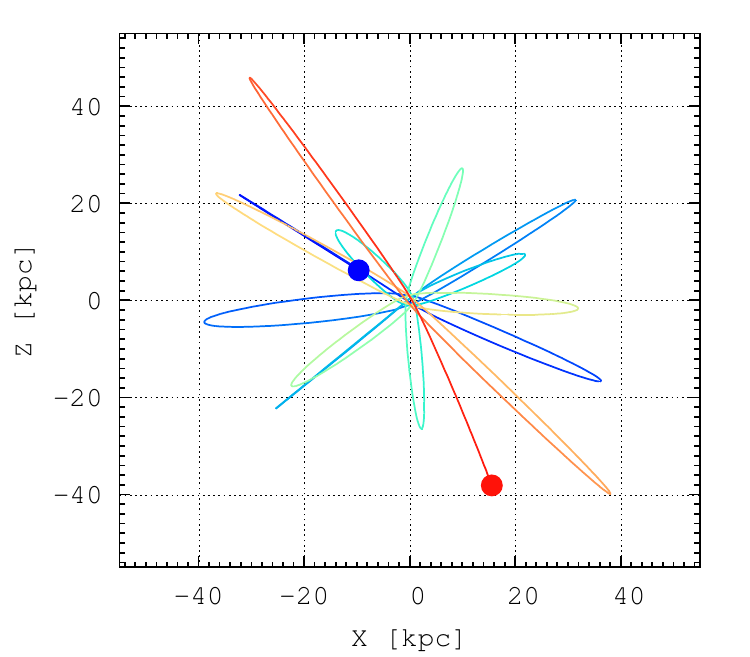}
\includegraphics[width=0.65\columnwidth]{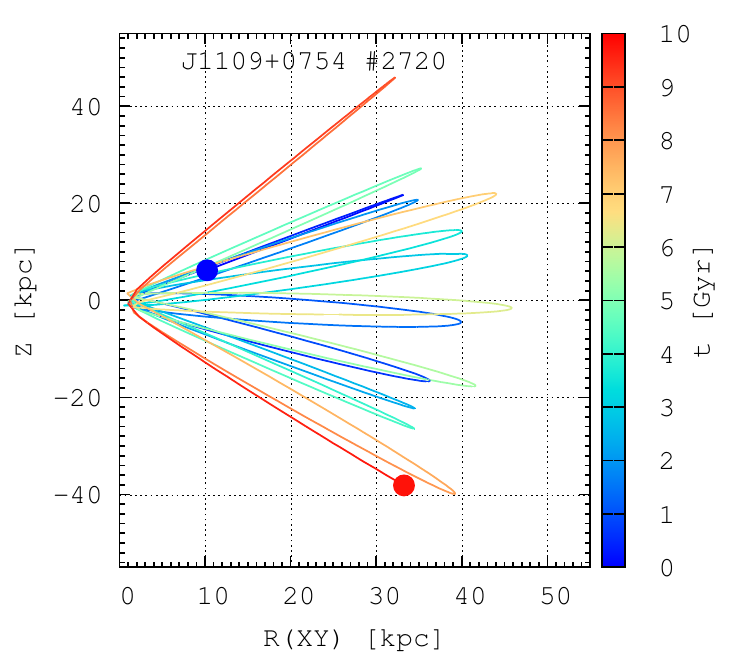}
\includegraphics[width=0.65\columnwidth]{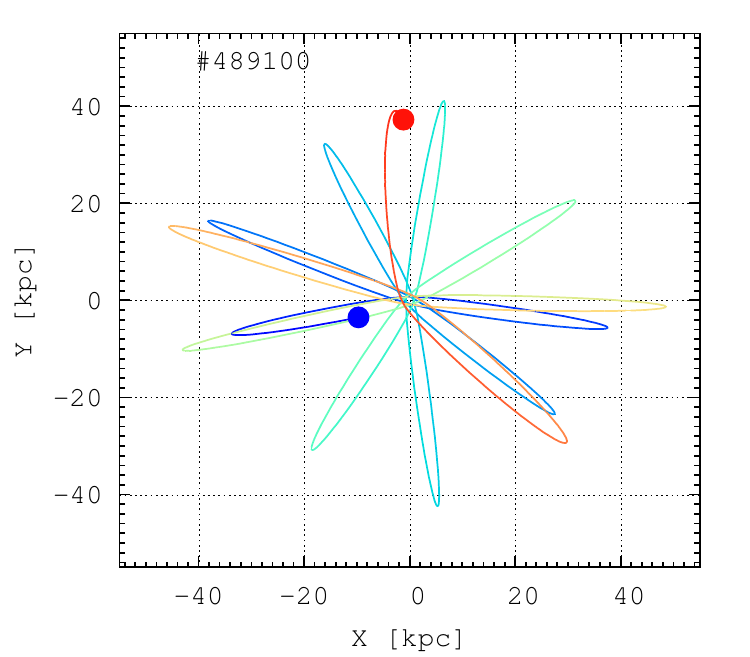}
\includegraphics[width=0.65\columnwidth]{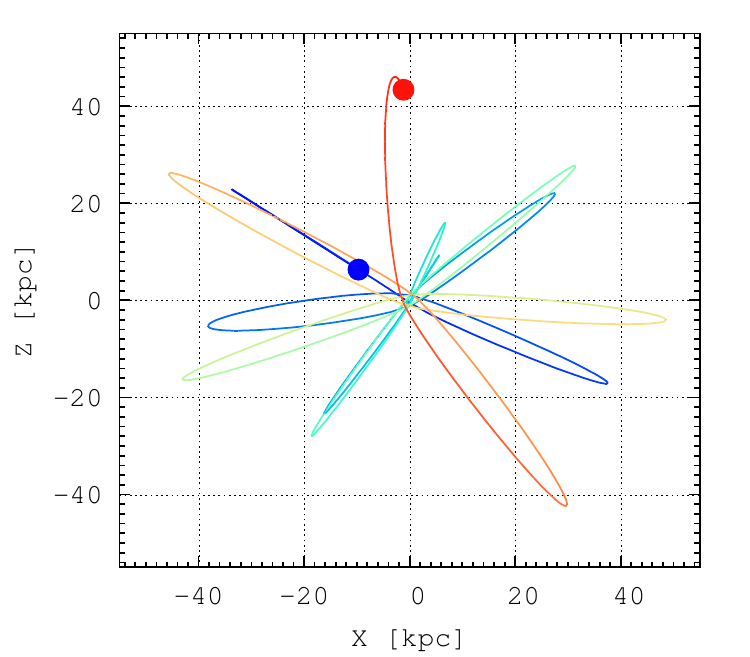}
\includegraphics[width=0.65\columnwidth]{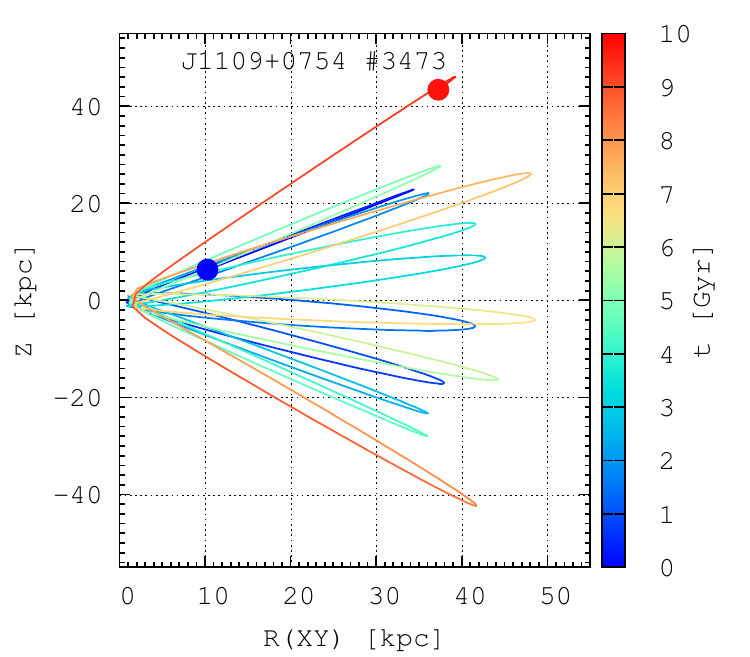}
\includegraphics[width=0.65\columnwidth]{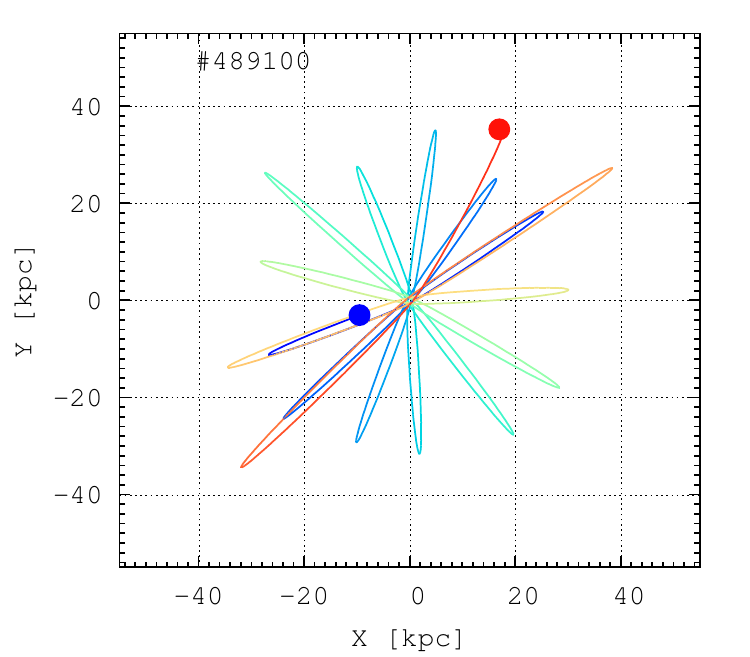}
\includegraphics[width=0.65\columnwidth]{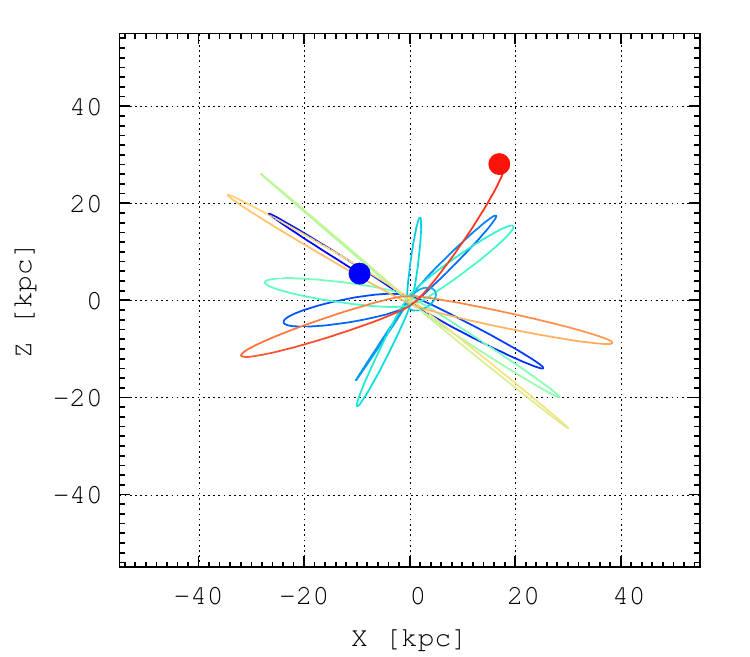}
\includegraphics[width=0.65\columnwidth]{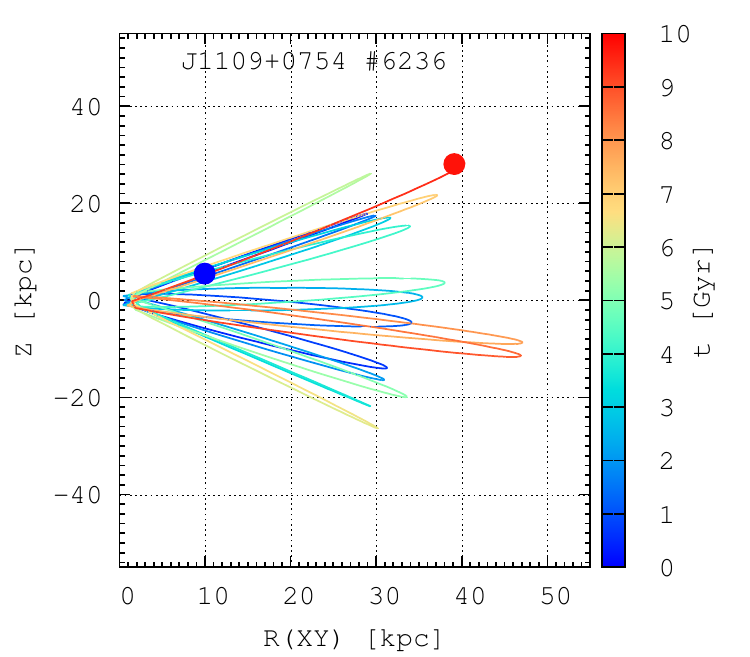}
\caption{Results of the backward integration of the orbital parameters of J1109+0754 in our selected \texttt{Illustris-TNG} halo {\tt \#489100}. 
Left panels show the X-Y coordinates of the orbits; the middle column presents the X-Z coordinates; the right panels show the time evolution in cylindrical coordinates R(XY)-Z. The color-coding corresponds to the time from 0 (i.e., the present) to 10\,Gyr backwards in time. Rows show different random realizations: {\tt \#2266}, {\tt \#2720}, {\tt \#3473}, {\tt \#6236} (inside $\pm~2\sigma$ from the initial 10,000 random values) of our star's initial positions and velocities.}
\label{fig:orbits}
\end{figure*}

Figure~\ref{fig:orbits} shows the backward orbits of a J1109+0754-like star, using the realizations that extend very far from the center, inside our selected subhalo {\tt \#489100}. The integration is performed in a backward fashion, from present day to a look-back time of 10\,Gyr. The orbits in Figure~\ref{fig:Orbits-static}, which are the result of a similar backward integration, {\it but in a static potential}, exhibits a more regular behavior, with the $\rapo$  and $\rperi$ distances being similar in different epochs. The difference between Figure~\ref{fig:orbits} and Figure~\ref{fig:Orbits-static} is particularly evident with our integration using subhalo \texttt{\#489100}. As can be seen from Figure~\ref{fig:orbits}, the $\rapo$ at early cosmic time is significantly larger than at present. Moreover, in some of the models, J1109+0754 enters the outer-halo region (up to $\sim$ 60\,kpc) after $\sim$ 6-7\,Gyrs. These orbits thus suggest a possible external origin of J1109+0754.

Collectively, the peculiar abundance pattern and the results from the the backward orbital integrations of J1109+0754 suggest that it was formed in a low-mass dwarf galaxy located $\sim$ 60\,kpc from the center of the Galaxy and accreted $\sim$ 6 - 7\,Gyr ago.

\section{Conclusions}\label{sec:conclusions}

In this study, we present a detailed chemo-dynamical analysis of the extremely metal-poor star LAMOST J1109+0754. The available radial-velocity measurements for this relatively bright ($V$=12.8) star suggests that J1109+0754 is not in a binary system, and thus its abundance pattern is not likely due to mass transfer from an unseen evolved companion; this is consistent with its observed sub-Solar carbon abundance (\abund{C}{Fe}= $-0.10$; natal abundance [C/Fe] = +0.66). In addition, J1109+0754 exhibits enhancements in the $\alpha$-elements ($\abund{\alpha}{Fe} \approx +0.4$), and large enhancements in the \textit{r}-process elements (\abund{Ba}{Fe}= $+0.25$ $\pm 0.21$ and \abund{Eu}{Fe}= = +0.94 $\pm 0.12$); indicating that J1109+0754 is a \cemp \textit{r}-II star, with no evidence of \textit{s}-process contribution (\abund{Ba}{Eu} = $-0.69$).

The observed light-element abundances do not deviate from the general trend observed for other metal-poor field stars reported in the literature. Moreover, the comparison between these abundances and the predicted yields of high-mass metal-free stars suggest a possible Population\,III progenitor with stellar mass of 22.5 M$_\odot$ and explosion energies 1.8-3.0 $10^{51}$ erg. The fitting result of this exercise supports the conclusion presented in \citet{Mardini_2019b}, which suggests that a stellar mass $\sim 20$\,M$_\odot$ progenitor may reflect the initial mass function of the first stars. Furthermore, it raises the question as to whether more-massive SNe might be more energetic, and therefore destroy their host halo and not allow for EMP star formation afterward.

The observed deviations of Sr, Y, and Zr from the scaled-Solar \textit{r}-process pattern indicate that the production of these elements (in the first \textit{r}-process peak) is likely to be different from the second and third \textit{r}-process peaks; these deviations are observed in other RPE stars. The universality of the main \textit{r}-process is confirmed for LAMOST J1109+0754 as well, due to the good agreement between the abundances of the elements with $Z > 56$ to the scaled-Solar \textit{r}-process residuals.

To carry out a detailed study of the possible orbital evolution of LAMOST J1109+0754, we carefully selected Milky Way-like galaxies from the \texttt{Illustris-TNG} simulation. We modeled the ORIENT from each candidate to be able to integrate the star's backward orbits. The results show that, in most cases, the star presents an extended Galactic orbit around 10\,Gyr ago (up to 60\,kpc, well into the outer-halo region), but is still bound to the Galaxy. These results, however, do not exclude the possibility that J1109+0754 has been accreted, as they are consistent with it being part of a dissolving dwarf galaxy positioned at about $\sim$ 60\,kpc from the center of the Galaxy $\sim$ 6 - 7\,Gyr ago. One caveat in our method to reconstruct the orbit is our selection criteria for Milky Way analog subhalos, specifically discarding those that exhibited "noisy" behavior, which may have biased the integration results toward orbits that remain bound. Another caveat is that backward integration may not be an ideal tool to establish the true phase-space coordinates of a star in the distant past, given the obvious limitations of the models for the gravitational potential at the very earliest times.

In future work, we plan to increase the number of Milky Way analog subhalos, and perform forward integration using initial conditions generated from the spatial and velocity distributions of particles in these subhalos. This will allow us to quantify the probability that LAMOST J1109+0754 has been accreted into the Milky Way, and answer questions such as when this accretion event may have occurred, and what was the likely mass of its progenitor subhalo.

\acknowledgements

\textbf{(Added: We thank an anonymous referee for positive remarks that helped us to improve this manuscript.)} M.K.M. thanks Ian Roederer and Tilman Hartwig for valuable discussions and helpful comments on earlier versions of the manuscript. This work was supported by the National Natural Science Foundation of China under grant Nos. 11988101 and 11890694, and National Key R\&D Program of China No.2019YFA0405502. VMP, TCB, and AF acknowledge partial support for this work from grant PHY 14-30152; Physics Frontier Center / JINA Center for the Evolution of the Elements (JINA-CEE), awarded by the US National Science Foundation. AF acknowledges support from NSF grant AST-1716251. MAS acknowledges financial support from the Alexander von Humboldt Foundation under the research program ``Black holes at all the scale''. MAS, MD, and MM acknowledge financial support by the Deutsche Forschungsgemeinschaft (DFG, German Research Foundation) -- Project-ID 138713538 -- SFB 881 (``The Milky Way System'', sub-projects A01, A02)." MAS, BA, and MM are grateful to Dylan Nelson, who helped in transferring the TNG dataset used in this work. The work of PB and MI was supported by the Deutsche Forschungsgemeinschaft (DFG, German Research Foundation) Project-ID 138713538, SFB 881 ("The Milky Way System"), and by the Volkswagen Foundation under the Trilateral Partnerships grant No. 97778. PB acknowledges support by the Chinese Academy of Sciences (CAS) through the Silk Road Project at NAOC, the President’s International Fellowship (PIFI) for Visiting Scientists program of CAS and the National Science Foundation of China (NSFC) under grant No. 11673032. MI acknowledges support by the National Academy of Sciences of Ukraine under the Young Scientists Grant No. 0119U102399. The work of PB was also partially supported under the special program of the National Academy of Sciences of Ukraine "Support for the development of priority fields of scientific research" (CPCEL 6541230). The  work  of BA was funded by a“Landesgraduiertenstipendium”of the University of Heidelberg and the Trilateral Collaboration Scheme project ”Accretion Processes in Galactic Nuclei” This work has made use of data from the European Space Agency (ESA) mission
{\it Gaia} (\url{https://www.cosmos.esa.int/gaia}), processed by the {\it Gaia}
Data Processing and Analysis Consortium (DPAC,
\url{https://www.cosmos.esa.int/web/gaia/dpac/consortium}). Funding for the DPAC
has been provided by national institutions, in particular the institutions
participating in the {\it Gaia} Multilateral Agreement.

\facility{
\texttt{APF},
\texttt{Gaia}, 
\texttt{Illustris-TNG:JupyterLab interface}~\citep{2019ComAC...6....2N}
}

\software{
\texttt{Astropy}~\citep{astropy:2013, astropy:2018}, 
\texttt{linemake}~(\url{https://github.com/vmplacco/linemake}),
\texttt{IRAF}~\citep{1986SPIE..627..733T, 1993ASPC...52..173T}, 
\texttt{Matplotlib}~\citep{hunter2007matplotlib}, 
\texttt{MOOG}~\citep{1973PhDT.......180S,2011AJ....141..175S},
\texttt{NumPy}~\citep{NumPyArray2011}, 
\texttt{\PGRAPE}~\citep{HGM2007},
\texttt{SciPy}~\citep{2020SciPy-NMeth},
\texttt{STARFIT}~\citep{2010ApJ...724..341H},
\texttt{TAME}~\citep{2015ascl.soft03003K}
} 

\newpage

\bibliography{new.ms}
\bibliographystyle{aasjournal}

\appendix \label{appendix}

\section{Parameter Fitting of the Milky-Way Analogs}

The bottom panels of Figure~\ref{fig:MW411321} show the extracted intrinsic parameters of the disk, as a function of the look-back time (red symbols) and the smoothing spline (blue dashed curve), respectively. The disk has three intrinsic parameters: length scales $a$ and $b$, and mass $M_\mathrm{d}$. It additionally has a direction vector and a center that needs to be calculated as well. Note that we did not include any bulge component as including it did not significantly improve the fitting of each snapshot.

The fitting procedure starts by searching for the density center of the stellar particles by recursively calculating the center of mass and removing star particles that lie outside of a predefined search radius. Then, the orientation of the disk is found by calculating the eigenvalues and eigenvectors of the quadrupole tensor of the particle distribution with respect to the density center. The largest eigenvalue corresponds to the shortest axis, and thus the orientation of the disk. As a consistency check, we also calculate the angular momentum of particles within the distribution's half-mass radius. We find that they agree very well (up to a sign because clockwise and counter-clockwise disks have the same quadrupole tensor) except for some snapshots, usually at high redshift, when no clear disk is formed, and the stellar particle's distribution is rather spherically symmetric.

The disk's mass is simply the sum of the stellar particle masses. The disk's length scales are calculated from the medians of the cylindrical radius and $z$ coordinates. The ratio between these medians is mapped to the ratio $b/a$ of the Miyamoto-Nagai disk model using a lookup table, and the scaling could be determined by either medians (we use the average of both). This method yields more robust results than a least-squares fitting of the density field of the particle distribution. Using the least-squares method often results in multiple solutions to $a$ and $b$, depending on the initial guesses. This numerical problem is a result of the stellar distribution not actually following a Miyamoto-Nagai disk model very closely. By considering the median values, outliers have a smaller effect.

Each halo has two intrinsic parameters: central density, $\rho_0$, and length scale, $a$. It additionally has a center that is calculated the same way as the disk's, but using the gas and dark matter particle distributions. The intrinsic parameters are calculated using a least-squares method, where the square differences between the cumulative mass from the center to that of the NFW model are minimized. The two top panels of Figure~\ref{fig:MW411321} show the extracted intrinsic parameters of the halo as a function of the look-back time (green symbols) and the smoothing spline fitting function (blue dashed curve).

We find that the offset between the disk and stellar halo centers did not exceed $\sim 1$\,kpc. This is typically much smaller than the stellar halo's length scale. Therefore, the calculations were done in a coordinate system centered at the density center of the disk component. The direction of the coordinate system was chosen in such a way that the disk pointed toward the positive $z$-direction in the last snapshot (representing the present day).

\section{Selected halo properties}

\begin{figure*}
\centering
\includegraphics{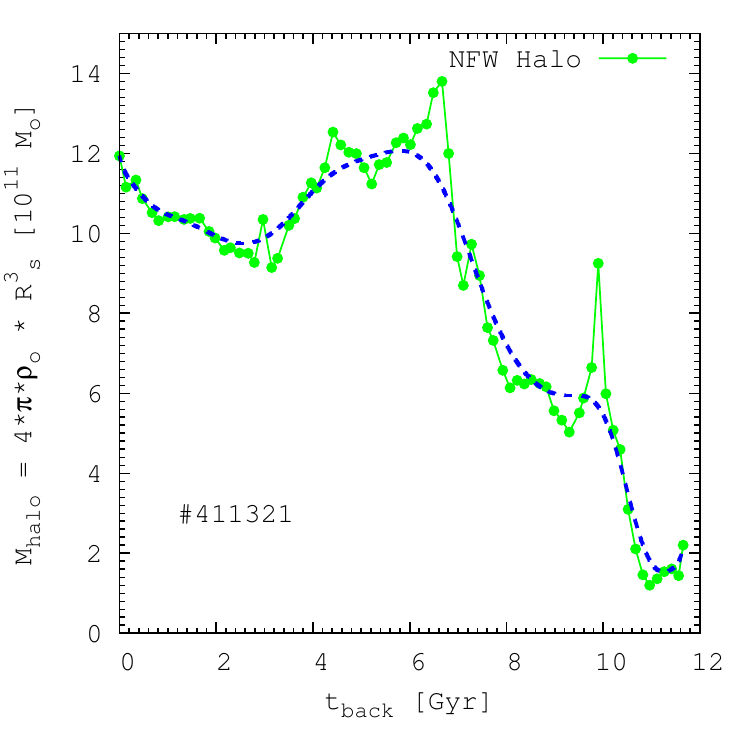}
\includegraphics{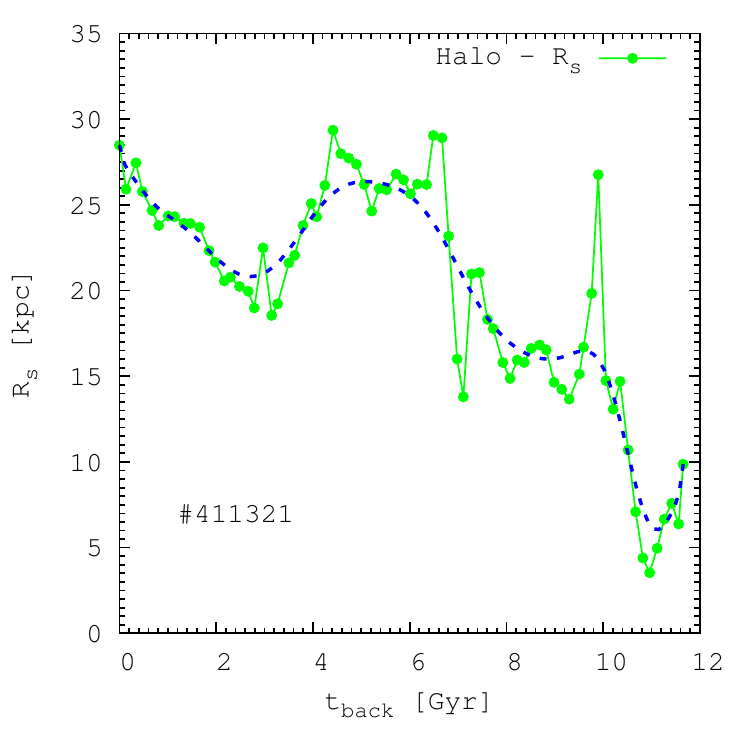}
\includegraphics{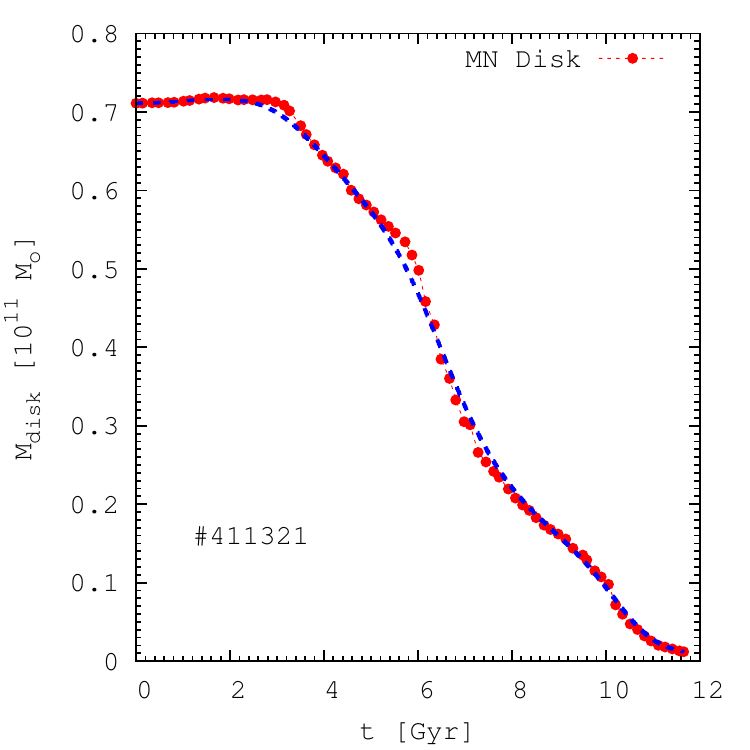}
\includegraphics{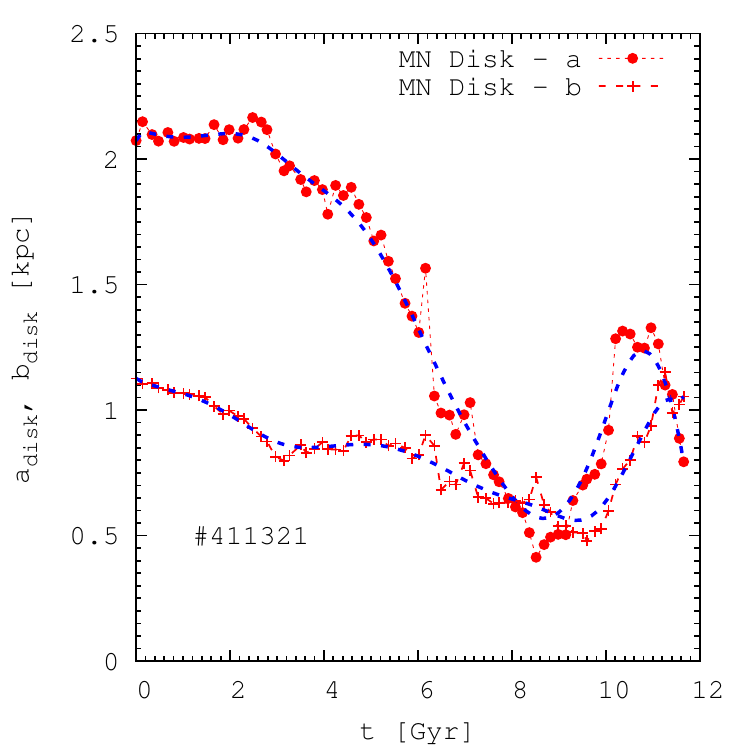}
\caption{Time evolution of sub-halo {\tt \#411321}. Halo and disk total 
masses are provided in units of 10$^{11}$\,M$_\odot$. The time axis is the look-back time in Gyr. We also present the NFW halo scale radius, R$_s$ (in kpc units), time evolution, together with the Miyamoto-Nagai disk scale parameters $a$ and $b$ (in kpc units). The points represent the direct data from the \texttt{Illustris-TNG} values (red or green 
points connected with red or green dashed lines). The dashed (blue) lines 
are the B\'ezier curve smooth data. For smoothing, we use 12,000 equally time-spaced data points.}
\label{fig:MW411321}
\end{figure*}

\begin{figure*}
\centering
\includegraphics{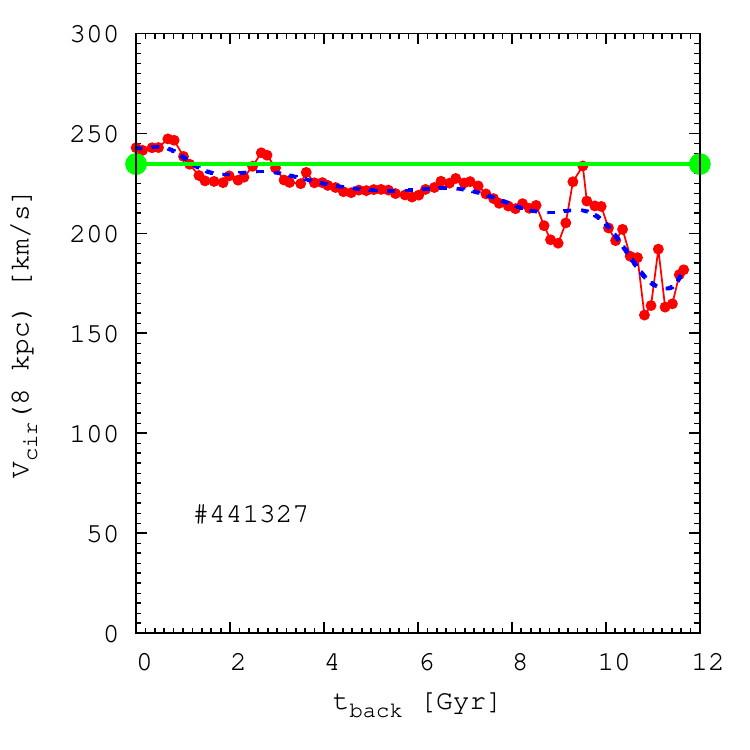}
\includegraphics{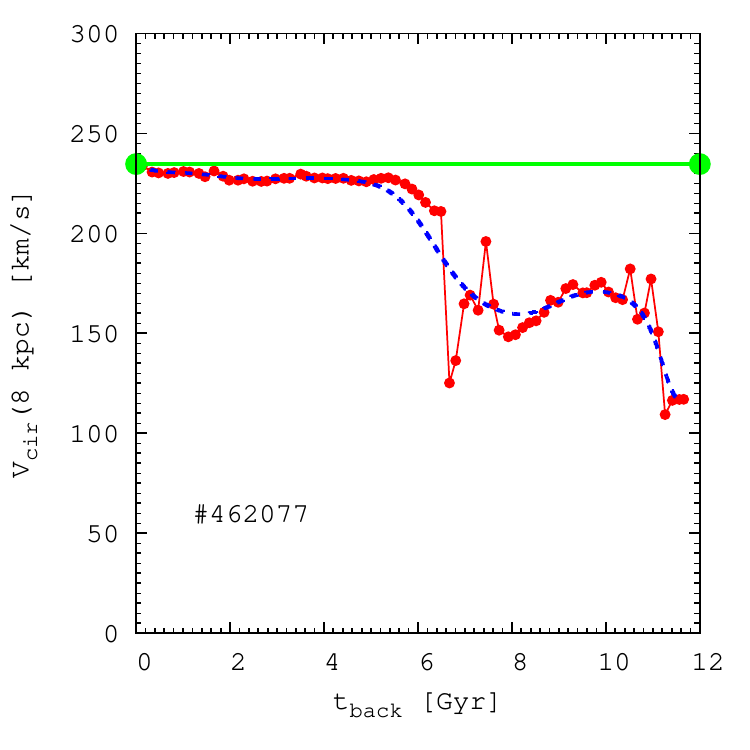}
\includegraphics{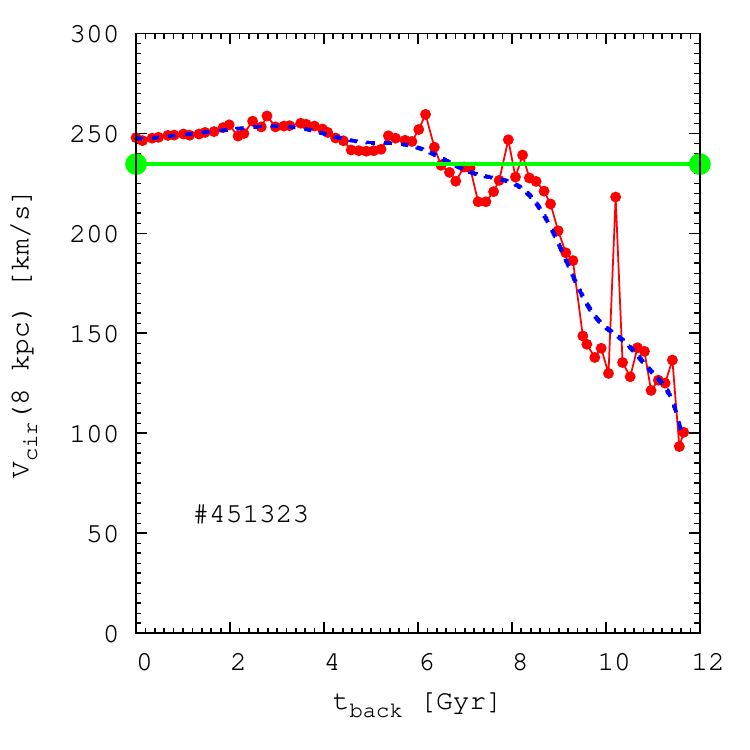}
\includegraphics{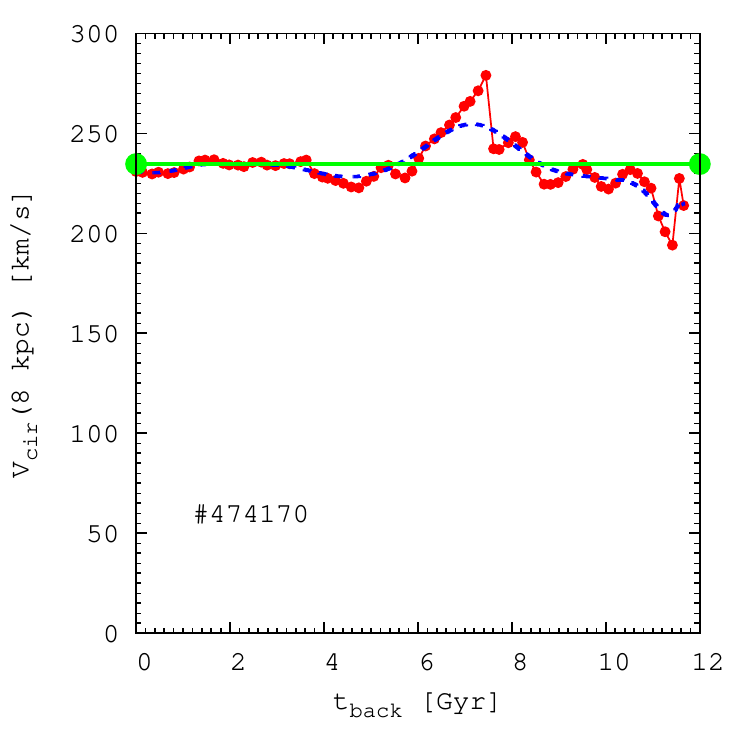}
\caption{Time evolution of the model circular rotational velocity at the  
location of the Sun (R$_{XY}$ = 8\,kpc, Z = 0) for selected sub-halos {\tt \#441327}, 
{\tt \#462077}, {\tt \#451323}, {\tt \#474170} (from left to right 
and from top to bottom). The horizontal lines (green) show our currently 
adopted circular velocity at the Sun's location of 234\,\mbox{km\,s$^{\rm -1}$}. The lines with red points 
show the actual data points from \texttt{Illustris-TNG}. 
The dashed blue lines show the smoothed data points.}
\label{fig:haloes}
\end{figure*}

Figure~\ref{fig:MW411321} denotes the mass- and size-parameter time evolution of one of our selected halos {\tt \#411321}\footnote{the number corresponds to the SubfindID at redshift $z=0$ in TNG100}. The NFW halo and the Miyamoto-Nagai disk total-mass time evolution in the units of 10$^{11}$\,M$_\odot$ are presented. The time axis corresponds to the look-back time in Gyr. We also present the NFW halo scale radius, R$_s$ (in kpc units), time evolution, together with the Miyamoto-Nagai disk scale parameters $a$ and $b$ (in kpc units). The connected points represent the direct \texttt{Illustris-TNG} values. For the real calculations, we smooth the data using the B\'ezier curve smoothing algorithm in the gnuplot software package. These fitted and smoothed curves (dashed lines) have one point for every 1\,Myr of time evolution. Integration timesteps could be significantly smaller than that, but the curves are smooth enough for simple linear interpolation between these points with sufficient accuracy. A high frequency is necessary in order to achieve a smooth integration of our stellar orbits in this time-dependent and complex external potential field.

Figure~\ref{fig:haloes} presents the time evolution of the circular rotational velocity at the location of the Sun (R$_{XY}$ = 8\,kpc, Z = 0) for the selected sub-halos ({\tt \#441327}, {\tt \#462077}, {\tt \#451323}, {\tt \#474170}). To obtain this value, we calculate for each time the enclosed total halo and disk mass inside the Solar cylinder. These curves show, in principle, the level of accuracy of our Milky Way approximation with these halos. It is clear that our selected \texttt{Illustris-TNG} halos agree reasonably well with the present day Solar Neighborhood rotation.
 \end{document}